\documentclass[12pt]{article}
\usepackage{epsfig}
      \def\di{\displaystyle}
      \def\bj{{\bf j}}
      \def\br{{\bf r}}
      \def\bp{{\bf p}}
      \def\bu{{\bf u}}
      \def\bs{{\bf s}}
      \def\Q{{\cal Q}}
      \def\R{{\cal R}}
      \def\P{{\cal P}}

      \def\L{{\cal L}}
      \def\N{{\cal N}}
      \def\J{{\cal J}}

\begin{document}

\vskip 5mm
\begin{center}
{\large\bf THE NUCLEAR SCISSORS MODE FROM
VARIOUS ASPECTS }\\
\vspace*{1cm}
{\large E.B. Balbutsev}\\
\vspace*{0.2cm}
{\it Joint Institute for Nuclear Research, 141980 Dubna, Moscow Region,
Russia}\\
\vspace*{0.5cm}
{\large P. Schuck}\\
\vspace*{0.2cm}
{\it Institut de Physique Nucleaire, Orsay Cedex 91406, France}
\end{center}

      \vspace{3cm}

\begin{abstract}
Three methods to describe collective motion, Random Phase
Approximation (RPA),
Wigner Function Moments (WFM) and the Green's Function (GF) method
are compared in detail and their physical content analyzed
 on an example of a simple model, the
harmonic oscillator with quadrupole--quadrupole residual interaction.
 It is
shown that they give identical formulae for eigenfrequencies and
transition probabilities of all collective excitations of the model,
including the scissors mode, which is the subject of our special
attention. The exact relation between the RPA and WFM variables
and the respective dynamical equations is established. The
transformation of the RPA spectrum into the one of WFM is explained.
The very close connection of the WFM method with the GF one is
demonstrated.
The normalization factor of the ``synthetic'' scissors state and its
overlap with physical states  are calculated analytically. The
orthogonality of the spurious state to all physical states is proved
rigorously. A differential equation describing the current
lines of RPA modes is established and the current lines of the
scissors mode analyzed as a superposition of rotational and
irrotational components.
\end{abstract}

\newpage

\section{Introduction}

The full analysis of the scissors mode in the framework of a solvable
model (harmonic oscillator with quadrupole--quadrupole residual
interaction (HO+QQ)) was
given in \cite{BaSc2}. Several points in the understanding of the
nature of this mode were clarified: for example, its
coexistence with the isovector giant quadrupole resonance (IVGQR),
the decisive role of the Fermi surface deformation, and several
things more.
The Wigner Function Moments (WFM) method was applied to derive
analytical expressions for currents of both coexisting modes, their
excitation energies,
magnetic and electric transition probabilities. Our formulae for
energies turned out to be identical with those derived by
Hamamoto and Nazarewicz \cite{Hamam} in the framework of the RPA.
However, little details are given in that reference of letter form.
In \cite{BalYaF} we investigated the relation between formulas for
transition probabilities derived by the two methods. It was shown
there, that also these formulas are identical. This coincidence
motivated us to undertake a systematic
comparison of the two approaches and to understand the connection and
differences between them. This is the goal of the present paper.
One of the important subjects of this comparison are the current
distributions. The WFM
method, a priori, can not give the exact results, because it
deals only with integrals over the whole phase space.
It would therefore be very interesting to evaluate
the accuracy of this approximation by comparing the results with
the currents obtained from RPA.
 Unfortunately, even for this simple model (HO+QQ) it is
impossible to derive in RPA the closed analytical expressions for
currents of the scissors mode and IVGQR. That is why we
consider in addition the Green's Function (GF) method, which allows one
to find explicit expressions for the currents directly.

 The HO+QQ model is a very convenient ground for this kind of
investigation, because all results can be obtained
analytically. There is no need to describe the merits
of the RPA or of the GF method -- they are very well known \cite{Ring}.
It is necessary, however, to say a few words about the WFM. Its idea
is based on the virial theorems of Chandrasekhar and Lebovitz
\cite{Chand}. Instead of writing the equations of motion for
microscopic amplitudes of particle hole excitations (RPA), one writes
the dynamical equations for various multipole phase space moments
of a nucleus. This allows one to achieve a more direct physical
interpretation of the studied phenomenon without going into its
detailed microscopic structure. The WFM method was successfully
applied to the study of
isoscalar and isovector giant multipole resonances and
low-lying collective modes of rotating and nonrotating nuclei with
various realistic forces \cite{Bal}. The results of WFM
were always very close to similar results obtained with the help of
RPA. In principle, this should be expected, because the basis of both
methods is the same: Time Dependent Hartree--Fock (TDHF) theory
in its small amplitude approximation. On the other hand it is
evident, that they are in general not equivalent, because the
truncation scheme is different. The detailed analysis
of the interplay of the two methods turns out to be useful also
from a ``practical" point of view: first and most importantly
it allows one to obtain
additional insight into the nature of the scissors mode; second
we find new exact mathematical results for the considered model.

The paper is organized as follows. In Section 2 we recall the
principal points of the WFM formalism and give a summary of the key
results of \cite{BaSc2} obtained by applying this method to the HO+QQ
model. The same model is considered in Section 3 in the frame of RPA:
the formulae for eigenfrequencies, electric and magnetic transition
probabilities of the scissors mode are derived, the ``synthetic"
scissors and spurious state are analyzed.
The exact interrelation between the RPA and WFM methods and between
their variables is established in Section 4. Section 5 is dedicated
to the GF method. The three methods are applied to derive analytical
formulae for lines of currents in Section 6. The mutual interplay
of the three methods is discussed in the conclusion. Various
mathematical details are given in Appendices A and B.

\section{The WFM method}

The basis of the method is the TDHF
equation for the one-body density matrix
$ \rho^{\tau}(\br_1, \br_2, t)=
\langle\br_1|\hat{\rho}^{\tau}(t)|\br_2\rangle $ :
\begin{equation}
\label{TDHF}
 i\hbar \frac{\partial \hat{\rho}^{\tau}}{\partial t}=
\left[ \hat{H}^{\tau} ,\hat{\rho}^{\tau}\right],
\end{equation}
where $ \hat{H}^{\tau} $ is the one-body self-consistent mean field
Hamiltonian depending implicitly on the density matrix and $\tau$ is
an isotopic spin index.
It is convenient to modify equation (\ref{TDHF}) introducing the
Wigner transform of the density matrix
\begin{equation}
\label{f}
 f^{\tau}(\br, \bp, t) = \int d^3s\: \exp(- i\bp
 \cdot \bs/\hbar)\rho^{\tau}(\br+\frac{\bs}{2}, \br-\frac{\bs}{2},t)
\end{equation}
and of the Hamiltonian
\begin{equation}
\label{Hw}
 H_W^{\tau}(\br,\bp)=\int d^3s\: \exp(-i\bp \cdot \bs/\hbar)
(\br+\frac{\bs}{2}\left|\hat{H}^{\tau}\right|\br-\frac{\bs}{2}).
\end{equation}
Using (\ref{f},\ref{Hw}) one arrives \cite{Ring} at

\begin{equation}
\label{fsin}
\frac{\partial f^{\tau}}{\partial t}=
\frac{2}{\hbar}\sin \left\{\frac{\hbar}{2}
\left[(\nabla)^H \cdot (\nabla^p)^f - (\nabla^p)^H \cdot
(\nabla)^f\right]\right\} H_W^{\tau} f^{\tau} ,
\end{equation}
 where the upper index on the bracket stands for the function
on which the operator in these brackets acts.
 It is shown in
\cite{Bal,BaSc}, that by integrating equation (\ref{fsin})
over the phase space $\{\bp,\br\}$ with the weights
$x_{i_1}x_{i_2}\ldots x_{i_k}p_{i_{k+1}}\ldots p_{i_{n-1}}p_{i_n}$,
where $k$ runs from $0$ to $n$, one can obtain a closed finite set
of dynamical equations for Cartesian tensors of the rank $n$.
Taking linear combinations of these equations one is able to represent
them through irreducible tensors, which play the role of collective
variables of the problem. However, it is more convenient to
derive the dynamical equations directly for irreducible tensors using
the technique of tensor products \cite{Varshal}. For this it is
necessary to
rewrite the Wigner function equation (\ref{fsin}) in terms of
cyclic variables
\begin{equation}
\label{sincyclic}
\frac{\partial f^{\tau}}{\partial t}=
\frac{2}{\hbar}\sin \left\{\frac{\hbar}{2}
\sum_{\alpha=-1}^1(-1)^{\alpha}
\left[(\nabla_{-\alpha})^H \cdot (\nabla^p_{\alpha})^f
- (\nabla^p_{-\alpha})^H \cdot (\nabla_{\alpha})^f
\right]\right\} H_W^{\tau} f^{\tau} ,
\end{equation}
 with $$\nabla_{+1}=-\frac{1}{\sqrt 2}(\frac{\partial}{\partial x_1}
+i\frac{\partial}{\partial x_2})~,\quad \nabla_0=\frac{\partial}
{\partial x_3}~,\quad
\nabla_{-1}=\frac{1}{\sqrt 2}(\frac{\partial}{\partial x_1}
-i\frac{\partial}{\partial x_2})~,$$
$$r_{+1}=-\frac{1}{\sqrt 2}(x_1+ix_2)~,\quad r_0=x_3~,\quad
r_{-1}=\frac{1}{\sqrt 2}(x_1-ix_2)$$and the analogous definitions
for $\nabla_{+1}^p~,\quad \nabla_{0}^p~,\quad \nabla_{-1}^p~, $ and
$p_{+1}~,\quad p_{0}~,\quad p_{-1}$.
The required equations are obtained by integrating
(\ref{sincyclic})
with different tensor products of $r_{\alpha}$ and $p_{\alpha}$.
Here we consider the case $n=2$.

\subsection{Model Hamiltonian, Equations of motion}

The microscopic Hamiltonian of the model, harmonic oscillator plus
separable quadrupole-quadrupole residual interaction is given by
\begin{eqnarray}
\label{Ham}
 H=\sum\limits_{i=1}^A(\frac{\hat\bp_i^2}{2m}+\frac{1}{2}m\omega^2\br_i^2)
+\bar{\kappa}
\sum_{\mu=-2}^{2}(-1)^{\mu}
 \sum\limits_i^Z \sum\limits_j^N
q_{2-\mu}(\br_i)q_{2\mu}(\br_j)
\nonumber\\
+\frac{1}{2}\kappa
\sum_{\mu=-2}^{2}(-1)^{\mu}
\{\sum\limits_{i\neq j}^{Z}
 q_{2-\mu}(\br_i)q_{2\mu}(\br_j)
+\sum\limits_{i\neq j}^{N}
 q_{2-\mu}(\br_i)q_{2\mu}(\br_j)\},
\end{eqnarray}
where the quadrupole operator $q_{2\mu}=\sqrt{16\pi/5}\,r^2Y_{2\mu}$
and $N,Z$ are the numbers of neutrons and protons, respectively.
 The mean field potential for protons (or neutrons) is
\begin{equation}
\label{poten}
V^{\tau}(\br,t)=\frac{1}{2}m\,\omega^2r^2+
\sum_{\mu=-2}^{2}(-1)^{\mu}\tilde Z_{2-\mu}^{\tau}(t)q_{2\mu}(\br),
\end{equation}
where $\tilde Z_{2\mu}^{\rm n}=\kappa Q_{2\mu}^{\rm n}
+\bar{\kappa}Q_{2\mu}^{\rm p}\,,\quad
\tilde Z_{2\mu}^{\rm p}=\kappa Q_{2\mu}^{\rm p}
+\bar{\kappa}Q_{2\mu}^{\rm n}\,$
and the quadrupole moments $Q_{2\mu}^{\tau}(t)$ are defined as
$$ Q_{2\mu}^{\tau}(t)=
\int\! d\{\bp,\br\}
q_{2\mu}(\br)f^{\tau}(\br,\bp,t)$$
 with
 $\int\! d\{\bp,\br\}\equiv
2 (2\pi\hbar)^{-3}\int\! d^3p\,\int\! d^3r$, where the factor 2
appears due to summation over spin degrees of freedom. To simplify
notation we omit spin indices, because we consider spin saturated
system without the spin--orbit interaction.

 Substituting spherical functions by tensor products
$\di r^2Y_{2\mu}=
\sqrt{\frac{15}{8\pi}}r_{2\mu}^2~,$
where
$$r^2_{\lambda\mu}\equiv
\{r\otimes r\}_{\lambda\mu}=\sum_{\sigma,\nu}
C_{1\sigma,1\nu}^{\lambda\mu}r_{\sigma}r_{\nu}$$
and $C_{1\sigma,1\nu}^{\lambda\mu}$ is the Clebsch-Gordan
coefficient,
 one has
\begin{equation}
\label{potenirr}
V^{\tau}=\frac{1}{2}m\,\omega^2r^2
+\sum_{\mu}(-1)^{\mu}Z_{2-\mu}^{\tau}r_{2\mu}^2.
\end{equation}
 Here
$$
Z_{2\mu}^{\rm n}=\chi R_{2\mu}^{\rm n}
+\bar{\chi}R_{2\mu}^{\rm p}\,,\quad
Z_{2\mu}^{\rm p}=\chi R_{2\mu}^{\rm p}
+\bar{\chi}R_{2\mu}^{\rm n}\,,\quad
\chi=6\kappa,\quad\bar\chi=6\bar\kappa,$$
\begin{equation}
\label{Rlmu}
 R_{\lambda\mu}^{\tau}(t)=
\int d\{\bp,\br\}
r_{\lambda\mu}^{2}f^{\tau}(\br,\bp,t).
\end{equation}

 Integration of equation (\ref{sincyclic}) with the weights
$r_{\lambda\mu}^2~,\,
(rp)_{\lambda\mu}\equiv\{r\otimes p\}_{\lambda\mu}$
and $p_{\lambda\mu}^2$
 yields the following set of equations \cite{BaSc2}:
\begin{eqnarray}
\label{quadr}
\frac{d}{dt}R_{\lambda\mu}^{\tau}
-\frac{2}{m}L^{\tau}_{\lambda\mu}&=&0,\quad \lambda=0,2
\nonumber\\
\frac{d}{dt}L^{\tau}_{\lambda\mu}
-\frac{1}{m}P_{\lambda\mu}^{\tau}+
m\,\omega^2R^{\tau}_{\lambda \mu}
-2\sqrt{5}\sum_{j=0}^2\sqrt{2j+1}\{_{2\lambda 1}^{11j}\}
(Z_2^{\tau}R_j^{\tau})_{\lambda \mu}
&=&0,
\quad \lambda=0,1,2
\nonumber\\
\frac{d}{dt}P_{\lambda\mu}^{\tau}
+2m\,\omega^2L^{\tau}_{\lambda \mu}
-4\sqrt{5}\sum_{j=0}^2\sqrt{2j+1}\{_{2\lambda 1}^{11j}\}
(Z_2^{\tau}L^{\tau}_j)_{\lambda \mu}
&=&0,\quad \lambda=0,2
\end{eqnarray}
 where $\{_{2\lambda 1}^{11j}\}$ is the Wigner
$6j$-symbol and the following
notation is introduced
$$P_{\lambda\mu}^{\tau}(t)=
\int\! d\{\bp,\br\}
p_{\lambda\mu}^{2}f^{\tau}(\br,\bp,t),\quad
 L_{\lambda\mu}^{\tau}(t)=
\int\! d\{\bp,\br\}
(rp)_{\lambda\mu}f^{\tau}(\br,\bp,t).$$
By definition
 $q_{2\mu}=\sqrt6r^2_{2\mu},\,
Q_{2\mu}^{\tau}=\sqrt6 R_{2\mu}^{\tau}$,
$R_{00}^{\tau}=-Q_{00}^{\tau}/\sqrt3$
with $Q_{00}^{\tau}=N_{\tau}\langle r^2\rangle$
being the mean square radius of
neutrons or protons. The tensor $L_{1\nu}^{\tau}$
is connected with angular momentum by the relations
$L_{10}^{\tau}=\frac{i}{\sqrt2}I_3^{\tau},\quad L_{1\pm 1}^{\tau}=
\frac{1}{2}(I_2^{\tau}\mp iI_1^{\tau}).$

  We rewrite equations (\ref{quadr}) in terms
 of the isoscalar and isovector variables
 $R_{\lambda\mu}=R_{\lambda\mu}^{\rm n}+R_{\lambda\mu}^{\rm p},\,
 \bar R_{\lambda\mu}=R_{\lambda\mu}^{\rm n}-R_{\lambda\mu}^{\rm p}$
 (and so on) with the isoscalar $\kappa_0=(\kappa+\bar{\kappa})/2$
 and isovector $\kappa_1=(\kappa-\bar{\kappa})/2$ strength constants.
 There is no problem to solve these equations numerically.
However, we want to simplify the situation as much as possible
to get the results in analytical form giving us a maximum of
insight into the nature of the modes.

1) We consider the problem in small-amplitude approximation.
 Writing all variables as a sum of their
equilibrium value plus a small deviation
$$R_{\lambda\mu}(t)=R_{\lambda\mu}^{eq}+\R_{\lambda\mu}(t),\quad
P_{\lambda\mu}(t)=P_{\lambda\mu}^{eq}+\P_{\lambda\mu}(t),\quad
L_{\lambda\mu}(t)=L_{\lambda\mu}^{eq}+\L_{\lambda\mu}(t),$$
$$\bar R_{\lambda\mu}(t)=\bar R_{\lambda\mu}^{eq}
+\bar \R_{\lambda\mu}(t),\quad
\bar P_{\lambda\mu}(t)=\bar P_{\lambda\mu}^{eq}
+\bar \P_{\lambda\mu}(t),\quad
\bar L_{\lambda\mu}(t)=\bar L_{\lambda\mu}^{eq}
+\bar \L_{\lambda\mu}(t),$$
we linearize the equations of motion in
$\R_{\lambda\mu},\,\P_{\lambda\mu},\,\L_{\lambda\mu}$ and
$\bar \R_{\lambda\mu},\,\bar \P_{\lambda\mu},\,\bar \L_{\lambda\mu}$.

2)We study non--rotating nuclei, i.e. nuclei with
$L_{1\nu}^{eq}=\bar L_{1\nu}^{eq}=0$.

3)Only axially symmetric nuclei with
$R_{2\pm2}^{eq}=R_{2\pm1}^{eq}=\bar R_{2\pm2}^{eq}=
\bar R_{2\pm1}^{eq}=0$ are considered.

4)Finally, we take
\begin{equation}
\label{Apr4}
\bar R_{20}^{eq}=\bar R_{00}^{eq}=0.
\end{equation}
This means that equilibrium deformation and mean square radius of
neutrons are supposed to be equal to that of protons.

Due to the approximation (\ref{Apr4}) the equations for isoscalar and
isovector systems are decoupled. Further, due to the axial symmetry
the angular momentum projection is a good quantum number. As a result,
every set of equations splits into five independent subsets with
$\mu=0,\pm 1,\pm 2.$ The detailed derivation of formulae for
eigenfrequencies and transition probabilities together with all
necessary explanations are given in \cite{BaSc2}. Here we write out
only the final results required for the comparison with respective
results obtained in the framework of RPA.

\subsection{Isoscalar eigenfrequencies}

The isoscalar subset of equations with $\mu=1$ is
\begin{eqnarray}
&&\dot \R_{21}-2\L_{21}/m=0,
\nonumber\\
&&\dot\L_{21}-\P_{21}/m+
\left[m\,\omega^2+ 2\kappa_0
(Q_{20}^{eq}+2Q_{00}^{eq})\right]\R_{21}=0,
\nonumber\\
&&\dot\P_{21}
+2[m\omega^2
+\kappa_0Q_{20}^{eq}]\L_{21}-6\kappa_0Q_{20}^{eq}\L_{11}
=0,
\nonumber\\
&&\dot \L_{11}=0.
\label{isosca1}
\end{eqnarray}
Using the self-consistent value of the strength constant
$\di{\kappa_0=-\frac{m\bar{\omega}^2}{4Q_{00}}}$,
(see Appendix~A) and the standard
definition of the deformation parameter
$\di Q_{20}=Q_{00}\frac{4}{3}\,\delta$ we reduce (\ref{isosca1}) to
\begin{eqnarray}
&&\dot \R_{21}-2\L_{21}/m=0,
\nonumber\\
&&\dot\L_{21}-\P_{21}/m=0,
\nonumber\\
&&\dot\P_{21}
+2m\bar\omega^2[(1+\frac{\delta}{3})\L_{21}+\delta\L_{11}]=0,
\nonumber\\
&&\dot \L_{11}=0.
\label{isosca2}
\end{eqnarray}
Imposing the time evolution via $\di{e^{-i\Omega t}}$ for all variables
one transforms (\ref{isosca2}) into a set of algebraic equations.
The eigenfrequencies are found from its characteristic equation
which reads
\begin{equation}
\label{haracis2}
\Omega^2[\Omega^2-2\bar{\omega}^2(1+\delta/3)]=0.
\end{equation}
The nontrivial solution of this equation gives the frequency of the
$\mu=1$ branch of the isoscalar GQR
\begin{equation}
\label{omegis}
\Omega^2=\Omega_{is}^2=2\bar{\omega}^2(1+\delta/3).
\end{equation}
Taking into account the relation (A.7) we find that this
result coincides with that of \cite{Suzuki}.
The trivial solution $\Omega=\Omega_0=0$ is characteristic of
nonvibrational mode corresponding to the obvious integral of motion
$\L_{11}=const$ responsible for the rotational degree of freedom.
This is usually called the `spurious' or `Goldstone' mode. It is easy
to find from (\ref{isosca2}) that $\L_{11}=0.$

\subsection{Isovector eigenfrequencies}

The information about the scissors mode is contained in the
subset of isovector equations
with $\mu=1$
\begin{eqnarray}
\label{scis}
&&\dot{\bar\R}_{21}-2\bar\L_{21}/m=0,
\nonumber\\
&&\dot{\bar\L}_{21}-\bar\P_{21}/m
+\left[m\,\omega^2+\kappa Q_{20}^{eq}
+4\kappa_1Q_{00}^{eq}\right]\bar\R_{21}=0,
\nonumber\\
&&\dot{\bar\P}_{21}
+2[m\omega^2+\kappa_0Q_{20}^{eq}]\bar\L_{21}
-6\kappa_0Q_{20}^{eq}\,\bar\L_{11}=0,
\nonumber\\
&&\dot{\bar\L}_{11}
+3\bar{\kappa}Q_{20}^{eq}\bar\R_{21}=0.
\end{eqnarray}
Supposing, as usual, the isovector constant $\kappa_1$ to be
proportional to the isoscalar one, $\kappa_1=\alpha\kappa_0$,
taking the self-consistent value for $\kappa_0$ and the standard
definition of $\delta$, we reduce (\ref{scis}) to
\begin{eqnarray}
\label{scis1}
&&\dot{\bar\R}_{21}-2\bar\L_{21}/m=0,
\nonumber\\
&&\dot{\bar\L}_{21}-\bar\P_{21}/m
+m\bar\omega^2(1-\alpha)(1+\frac{\delta}{3})\bar\R_{21}=0,
\nonumber\\
&&\dot{\bar\P}_{21}
+2m\bar\omega^2[(1+\frac{\delta}{3})\bar\L_{21}+\delta\bar\L_{11}]=0,
\nonumber\\
&&\dot{\bar\L}_{11}
-m\bar\omega^2\delta(1-\alpha)\R_{21}=0.
\end{eqnarray}
Imposing the time evolution via $\di{e^{-i\Omega t}}$
one transforms (\ref{scis1}) into a set of algebraic equations
with the characteristic equation
\begin{equation}
\label{harac2}
\Omega^4-2\Omega^2\bar{\omega}^2(2-\alpha)(1+\delta/3)
+4\bar{\omega}^4(1-\alpha)\delta^2=0.
\end{equation}
Its solutions are
\begin{equation}
\label{Omeg2}
\Omega^2_{\pm}=\bar{\omega}^2(2-\alpha)(1+\delta/3)
\pm \sqrt{\bar{\omega}^4(2-\alpha)^2(1+\delta/3)^2
-4\bar{\omega}^4(1-\alpha)\delta^2}.
\end{equation}
The high-lying solution $\Omega_+$ gives the frequency $\Omega_{iv}$
of the $\mu=1$ branch of the isovector GQR.
The low-lying solution $\Omega_-$ gives the frequency $\Omega_{sc}$
of the scissors mode.

 We adjust $\alpha$ from the fact that the IVGQR is
experimentally known to lie practically at twice the energy of the
isoscalar GQR. In our model the experimental situation is satisfied
by $\alpha=-2$. Then
\begin{eqnarray}
\Omega^2_{iv}=4\bar\omega^2
\left(1+\frac{\delta}{3}+\sqrt{(1+\frac{\delta}{3})^2-
 \frac{3}{4}\delta^2}\,\right),
\quad
\Omega^2_{sc}=4\bar\omega^2
\left(1+\frac{\delta}{3}-\sqrt{(1+\frac{\delta}{3})^2-
\frac{3}{4}\delta^2}\,\right).
\label{Omeg2fin}
\end{eqnarray}

The scissors mode energy in the limit of small deformation is
\begin{equation}
E_{sc}\approx\sqrt{\frac{3}{2}}\,\hbar\omega\,\delta,
\end{equation}
which is quite close to the result of Hilton \cite{Hilt92}:
$E_{sc}\approx\sqrt{1+0.66}\,\hbar\omega\,\delta.$
Taking $\hbar\omega=45.2/A^{1/3}$ MeV (what corresponds to
$r_0=1.15$ fm used in \cite{Lipp}), one obtains
$$E_{sc}\approx55.4\,\delta A^{-1/3}\,{\rm MeV},$$
which practically coincides with the result of Lipparini and Stringari
\cite{Lipp} (formula (18): $E_{sc}\simeq56\,\delta A^{-1/3}$ MeV)
obtained with the help of a microscopic approach based on the
evaluation of sum rules.
 Both results are not very far from the experimental \cite{Richter}
value: $E_{sc}\approx 66\delta A^{-1/3}$ MeV.

It is interesting to study the role of the Fermi surface deformation
for the formation of IVGQR and the scissors mode.
Neglecting the variable $\bar \P_{21}(t)$ in (\ref{scis1}) we find that
the frequency of IVGQR (being determined mainly by the neutron-proton
interaction) is changed not very much:
$$\Omega_{iv}^2=2\bar\omega^2(1-\alpha)(1+\delta/3).$$
Comparing this formula (for $\alpha=-2$ ) with (\ref{Omeg2fin}) one sees,
that in the limit of small deformation one obtains
 $\Omega_{iv}^2\simeq6\omega^2_0$ instead of
$\Omega_{iv}^2\simeq8\omega^2_0$. One should recall that also for the
Isovector Giant Dipole Resonance
the distortion of Fermi sphere plays only a minor role.

 It is also easy to see
that omitting $\bar \P_{21}(t)$ in (\ref{scis1}), one
obtains zero energy for the scissors mode independent of the
strength of the residual interaction.
Thus, the nuclear elasticity
discovered by G.F.Bertsch \cite{Bertsch} is the single origin for the
restoring force of the scissors mode.
 So one can conclude that this mode is in its essence a pure quantum
mechanical phenomenon. This agrees with the conclusion of the papers
\cite{Lipp9,ZawSp}: classically (i.e., without Fermi surface
deformation) the scissors mode is a zero energy mode.

\subsection{Linear response and transition probabilities}

A direct way of calculating the reduced transition probabilities
is provided by the theory of the linear response of a system to a
weak external field
$$\hat F(t)=\hat F\,{\rm exp}(-i\Omega t)+
\hat F^{\dagger}\,{\rm exp}(i\Omega t),$$
where $\hat F=\sum_{s=1}^A\hat f_s$ is a one-body operator.
A convenient form of the response theory is e.g. given by Lane
\cite{Lan} (see also section 4).
The matrix elements of the operator $\hat F$ obey the relation
\begin{equation}
\label{Fmatel}
|<\nu|\hat F|0>|^2=
\hbar\lim_{\Omega\to\Omega_{\nu}}(\Omega-\Omega_{\nu})
\overline{<\psi|\hat F|\psi>\exp(-i\Omega t)},
\end{equation}
where $|0>$ and $|\nu>$ are the stationary wave functions of the
ground and unperturbed excited states; $\psi$ is the perturbed
wavefunction of the ground state, $\Omega_{\nu}=(E_{\nu}-E_0)/\hbar$
are the
normal frequencies, the bar means averaging over a time interval much
larger than $1/\Omega$, $\Omega$ being the frequency of the external
field $\hat F(t)$.

{\bf Electric excitations} are described by the operator
\begin{equation}
\label{Oelec}
\hat F=\hat F_{2\mu}^{\rm p}=\sum_{s=1}^Z\hat f_{2\mu}(s), \quad
\hat f_{2\mu}=e\,r^2Y_{2\mu}=\beta r^2_{2\mu},\quad
\beta=e\sqrt{\frac{15}{8\pi}}
\end{equation}
whose expectation value (in accordance with \cite{BaSc2}) is
$$<\psi|\hat F_{2\mu}^{\rm p}|\psi>=\beta R_{2\mu}^{\rm p}=
\frac{1}{2}\beta (R_{2\mu}-\bar R_{2\mu}).$$
Transition probabilities are given \cite{BaSc2} by the following
formulae
\begin{eqnarray}
\label{E2sc}
B(E2)_{sc}=2|<sc|\hat F_{21}^{\rm p}|0>|^2
=\frac{e^2\hbar}{m}\frac{5}{8\pi}Q_{00}
\frac{(1+\delta/3)\Omega_{sc}^2-2(\bar\omega\delta)^2}
{\Omega_{sc}(\Omega^2_{sc}-\Omega_{iv}^2)}.
\end{eqnarray}

\begin{eqnarray}
\label{E2iv}
B(E2)_{iv}=2|<iv|\hat F_{21}^{\rm p}|0>|^2
=\frac{e^2\hbar}{m}\frac{5}{8\pi}Q_{00}
\frac{(1+\delta/3)\Omega_{iv}^2-2(\bar\omega\delta)^2}
{\Omega_{iv}(\Omega^2_{iv}-\Omega_{sc}^2)}.
\end{eqnarray}

\begin{eqnarray}
\label{E2is}
B(E2)_{is}=2|<is|\hat F_{21}^{\rm p}|0>|^2
=\frac{e^2\hbar}{m}\frac{5}{8\pi}Q_{00}
[(1+\delta/3)\Omega_{is}^2
-2(\bar\omega\delta)^2]/[\Omega_{is}]^3.
\end{eqnarray}
These three formulae can be joined into one expression
by a simple transformation of the denominators. Really,
we have from (\ref{Omeg2})
\begin{eqnarray}
\pm(\Omega^2_{iv}-\Omega_{sc}^2)&=&\pm(\Omega^2_{+}-\Omega_{-}^2)=
\pm 2 \sqrt{\bar{\omega}^4(2-\alpha)^2(1+\delta/3)^2
-4\bar{\omega}^4(1-\alpha)\delta^2}
\nonumber\\
&=&2\Omega^2_{\pm}-2\bar{\omega}^2(2-\alpha)(1+\delta/3)=
2\Omega^2_{\pm}-(2-\alpha)(\omega_x^2+\omega_z^2).
\label{Deno}
\end{eqnarray}
Using these relations in formulae (\ref{E2sc}) and
(\ref{E2iv}) we
obtain the expression for the $B(E2)$ values valid for all three
excitations
\begin{eqnarray}
\label{BE2gen}
B(E2)_{\nu}=2|<\nu|\hat F_{21}^{\rm p}|0>|^2
=\frac{e^2\hbar}{m}\frac{5}{16\pi}Q_{00}
\frac{(1+\delta/3)\Omega_{\nu}^2-2(\bar\omega\delta)^2}
{\Omega_{\nu}[\Omega^2_{\nu}-\bar{\omega}^2(2-\alpha)(1+\delta/3)]}.
\end{eqnarray}
The isoscalar value (\ref{E2is}) is obtained by assuming $\alpha=1.$

{\bf Magnetic excitations}
are described by the operator
\begin{equation}
\label{Omagn}
\hat F=\hat F_{1\mu}^{\rm p}=\sum_{s=1}^Z\hat f_{1\mu}(s), \quad
\hat f_{1\mu}=-i\nabla
(rY_{1\mu})\cdot[\br\times\nabla]\mu_N=\gamma(r\hat p)_{1\mu},\quad
\mu_N=\frac{e\hbar}{2mc}.
\end{equation}
Its expectation value was calculated in \cite{BaSc2} to be
$$<\psi|\hat F_{1\mu}^{\rm p}|\psi>=\gamma L_{1\mu}^{\rm p}=
\frac{\gamma}{2}(L_{1\mu}-\bar L_{1\mu})=
\frac{\gamma}{2}(\L_{1\mu}-\bar \L_{1\mu}),\quad
\gamma=-i\frac{e}{2mc}\sqrt{\frac{3}{2\pi}}.$$
Then one finds the following expessions \cite{BaSc2} for transition
probabilities
\begin{eqnarray}
\label{scimat}
B(M1)_{sc}=2|<sc|\hat F_{11}^{\rm p}|0>|^2=
\frac{1-\alpha}{4\pi}\frac{m\bar\omega^2}{\hbar}
Q_{00}\delta^2\frac{\Omega_{sc}^2-2(1+\delta/3)\bar\omega^2}
{\Omega_{sc}(\Omega^2_{sc}-\Omega_{iv}^2)}\,\mu_N^2,
\end{eqnarray}
\begin{eqnarray}
\label{M1iv}
B(M1)_{iv}=2|<iv|\hat F_{11}^{\rm p}|0>|^2
=\frac{1-\alpha}{4\pi}\frac{m\bar\omega^2}{\hbar}Q_{00}
\delta^2\frac{\Omega_{iv}^2-2(1+\delta/3)\bar\omega^2}
{\Omega_{iv}(\Omega^2_{iv}-\Omega_{sc}^2)}\,\mu_N^2.
\end{eqnarray}
Using relations (\ref{Deno}) in formulae (\ref{scimat}) and
(\ref{M1iv}), we
obtain the expression for the $B(M1)$ values valid for both excitations
\begin{eqnarray}
\label{BM1gen}
B(M1)_{\nu}=2|<\nu|\hat F_{11}^{\rm p}|0>|^2
=\frac{1-\alpha}{8\pi}\frac{m\bar\omega^2}{\hbar}Q_{00}\delta^2
\frac{\Omega_{\nu}^2-2(1+\delta/3)\bar\omega^2}
{\Omega_{\nu}[\Omega^2_{\nu}-\bar{\omega}^2(2-\alpha)(1+\delta/3)]}
\,\mu_N^2.
\end{eqnarray}

 Taking into account the relation
${\di Q_{00}^0\frac{m\omega_0}{\hbar}
\simeq\frac{1}{2}\left(\frac{3}{2}A\right)^{4/3}}$, which is usually
\cite{Solov} used to fix the value of the harmonic oscillator
frequency $\omega_0$ , we obtain the following estimate for the
transition probability of the scissors mode:
$$B(M1)\!\uparrow=2\,|<sc|\hat F_{11}^{\rm p}|0>|^2=
\frac{(3/2)^{11/6}}{16\pi}A^{4/3}\delta\,\mu_N^2
=0.042A^{4/3}\delta\,\mu_N^2,$$
which practically coincides with the result of
\cite{Lipp}: $B(M1)\!\uparrow=0.043A^{4/3}\delta\,\mu_N^2$,
obtained with  the help of the microscopic approach based on the
evaluation of the sum rules.

Concluding this section one should mention, that all magnetic
and electric modes of the considered model satisfy the energy weighted
sum rule \cite{BaSc2}, the contribution of the spurious (or
Goldstone) mode being nonzero.
It is interesting to compare the contributions of the scissors mode
and the spurious mode. The scissors mode (for small $\delta$) yields:
\begin{eqnarray}
2E_{sc}|<sc|\hat F_{21}^{\rm p}|0>|^2
\simeq\frac{5}{128\pi}e^2\frac{\hbar^2}{m}Q_{00}\delta^2.
\label{sumsc}
\end{eqnarray}
The spurious mode yields:
\begin{eqnarray}
2\hbar\Omega_0|<\Omega_0|\hat F_{21}^{\rm p}|0>|^2=
\frac{5}{8\pi}e^2\frac{\hbar^2}{m}Q_{00}\frac{\delta^2}{1+\delta/3}.
\label{sumnonv}
\end{eqnarray}
It is seen that the contribution of the spurious mode is
approximately 16 times larger than the one of the scissors mode.
This is a very significant number demonstrating the importance of
excluding the spurious state from the theoretical results. Indeed,
to describe correctly such a subtle phenomenon as the scissors mode,
it is compulsory to eliminate the errors from spurious motion whose
value can be an order of magnitude larger than the phenomenon under
consideration.

\section{Random Phase Approximation (RPA)}

In this section we now want to derive the analogous equations for
energies and transition probabilities from
standard RPA theory. RPA equations in the notation of \cite{Ring} are
\begin{eqnarray}
\label{RPA}
\sum_{n,j}\left\{\left[\delta_{ij}\delta_{mn}(\epsilon_m-
\epsilon_i)+\bar v_{mjin}\right]X_{nj}+\bar v_{mnij}Y_{nj}\right\}
=\hbar\Omega X_{mi},
\nonumber\\
\sum_{n,j}\left\{\bar v_{ijmn}X_{nj}+
\left[\delta_{ij}\delta_{mn}(\epsilon_m-
\epsilon_i)+\bar v_{inmj}\right]Y_{nj}\right\}
=-\hbar\Omega Y_{mi}.
\end{eqnarray}
According to the definition of the schematic model in \cite{Ring},
the matrix elements of the residual interaction corresponding to the
Hamiltonian (\ref{Ham}) are written as
$$\bar v_{mjin}=
\kappa_{\tau\tau'}\Q_{im}^{\tau*}\Q_{jn}^{\tau'}$$
with $\Q_{im}\equiv<i|q_{21}|m>$ and $\kappa_{\rm nn}
=\kappa_{\rm pp}=\kappa,\quad \kappa_{\rm np}=\bar\kappa$.
This interaction distinguishes between protons and neutrons, so we have
to introduce the isospin indices $\tau,\,\tau'$
 into the set of RPA equations
(\ref{RPA}):
\begin{eqnarray}
\label{DD}
(\epsilon_m^{\tau}-\epsilon_i^{\tau})X_{mi}^{\tau}+
\sum_{n,j,\tau'}\kappa_{\tau\tau'}\Q_{im}^{\tau*}\Q_{jn}^{\tau'}
X_{nj}^{\tau'}+
\sum_{n,j,\tau'}\kappa_{\tau\tau'}\Q_{im}^{\tau*}\Q_{nj}^{\tau'}
Y_{nj}^{\tau'}
=\hbar\Omega X_{mi}^{\tau},
\nonumber\\
\sum_{n,j,\tau'}\kappa_{\tau\tau'}\Q_{mi}^{\tau*}\Q_{jn}^{\tau'}
X_{nj}^{\tau'}+
(\epsilon_m^{\tau}-\epsilon_i^{\tau})Y_{mi}^{\tau}+
\sum_{n,j,\tau'}\kappa_{\tau\tau'}\Q_{mi}^{\tau*}\Q_{nj}^{\tau'}
Y_{nj}^{\tau'}
=-\hbar\Omega Y_{mi}^{\tau}.
\end{eqnarray}
 The solution of these equations is
\begin{equation}
X_{mi}^{\tau}=
\frac{\Q_{im}^{\tau*}}{\hbar\Omega-\epsilon_{mi}^{\tau}}K^{\tau},
\quad
Y_{mi}^{\tau}=
-\frac{\Q_{mi}^{\tau*}}{\hbar\Omega+\epsilon_{mi}^{\tau}}K^{\tau}
\label{XY}
\end{equation}
with $\epsilon_{mi}^{\tau}=\epsilon_m^{\tau}-\epsilon_i^{\tau}$ and
$K^{\tau}=\sum_{\tau'}\kappa_{\tau\tau'}C^{\tau'}.$

The constant $C^{\tau}$ is defined as
$C^{\tau}=\sum_{n,j}(\Q_{jn}^{\tau}X_{nj}^{\tau}
+\Q_{nj}^{\tau}Y_{nj}^{\tau}).$
Using here the expressions for $X_{nj}^{\tau}$ and
$Y_{nj}^{\tau}$ given above,
one derives the useful relation
\begin{equation}
C^{\tau}=2S^{\tau}K^{\tau}
=2S^{\tau}\sum_{\tau'}\kappa_{\tau\tau'}C^{\tau'},
\label{CSK}
\end{equation}
where the following notation is introduced:
\begin{equation}
S^{\tau}=\sum_{mi}|\Q_{mi}^{\tau}|^2\frac{\epsilon_{mi}^{\tau}}
{E^2-(\epsilon_{mi}^{\tau})^2}
\label{S}
\end{equation}
with $E=\hbar\Omega$.
Let us write out the relation (\ref{CSK}) in detail
\begin{eqnarray}
C^{\rm n}-2S^{\rm n}(\kappa C^{\rm n}+\bar\kappa C^{\rm p})=0,
\nonumber\\
C^{\rm p}-2S^{\rm p}(\bar\kappa C^{\rm n}+\kappa C^{\rm p})=0.
\label{Cnp}
\end{eqnarray}
The condition for existence of a nontrivial solution of this set of
equations gives the secular equation
\begin{equation}
(1-2S^{\rm n}\kappa)(1-2S^{\rm p}\kappa)
-4S^{\rm n}S^{\rm p}\bar\kappa^2=0.
\label{Secul1}
\end{equation}
Making obvious linear combinations of the two equations in
(\ref{Cnp}), we write them in terms of isoscalar and
isovector constants $C=C^{\rm n}+C^{\rm p},\,\bar C=C^{\rm n}-C^{\rm p}$
\begin{eqnarray}
C-2(S^{\rm n}+S^{\rm p})\kappa_0 C
-2(S^{\rm n}-S^{\rm p})\kappa_1 \bar C=0,
\nonumber\\
\bar C-2(S^{\rm n}-S^{\rm p})\kappa_0 C
-2(S^{\rm n}+S^{\rm p})\kappa_1 \bar C=0.
\label{Cnpiso}
\end{eqnarray}
 Approximation (\ref{Apr4}) allows us to decouple the equations for
isoscalar and isovector constants. Really, in this case
$S^{\rm n}=S^{\rm p}\equiv S/2$;
hence, we obtain two secular equations
\begin{equation}
1-2S\kappa_0=0,
 \quad \mbox{or}\quad 1-S\kappa=S\bar\kappa
\label{Secis}
\end{equation}
in the isoscalar case and
\begin{equation}
1-2S\kappa_1=0,
 \quad \mbox{or}\quad 1-S\kappa=-S\bar\kappa
\label{Seciv}
\end{equation}
in the isovector one, the difference of both lies in the
strength constants only. Having in mind the relation
$\kappa_1=\alpha \kappa_0$, we come to the conclusion
that it is sufficient to analyze the isovector case only -- the
results for isoscalar one are obtained by assuming $\alpha=1$.

\subsection{Eigenfrequencies}

The detailed expression for the isovector secular equation is
\begin{equation}
\frac{1}{2\kappa_1}=
\sum_{mi}|\Q_{mi}|^2\frac{\epsilon_{mi}}
{E^2-\epsilon_{mi}^2}.
\label{Secular}
\end{equation}
The operator $\Q$ has only two types of
nonzero matrix elements $\Q_{mi}$ in the deformed oscillator basis.
Matrix elements of the first type couple states of the same major
shell. All corresponding transition energies are degenerate:
$\epsilon_m-\epsilon_i=\hbar(\omega_x-\omega_z)\equiv\epsilon_0$.
Matrix elements of the second type couple states of the different
major shells with $\Delta N=2$. All corresponding transition energies
are degenerate too:
$\epsilon_m-\epsilon_i=\hbar(\omega_x+\omega_z)\equiv\epsilon_2$.
Therefore, the secular equation can be rewritten as
\begin{equation}
\frac{1}{2\kappa_1}=\frac{\epsilon_0\Q_0}{E^2-\epsilon_0^2}+
\frac{\epsilon_2\Q_2}{E^2-\epsilon_2^2}.
\label{Secul}
\end{equation}
The sums $\di \Q_0=\sum_{mi(\Delta N=0)}|\Q_{mi}|^2$ and
$\di \Q_2=\sum_{mi(\Delta N=2)}|\Q_{mi}|^2$ can be calculated
analytically (see Appendix B):
\begin{equation}
\label{sums}
\Q_0=\frac{Q_{00}}{m\bar\omega^2}\epsilon_0,
\quad
\Q_2=\frac{Q_{00}}{m\bar\omega^2}\epsilon_2.
\end{equation}
Let us transform the secular equation (\ref{Secul}) in polynomial form
$$E^4-E^2[(\epsilon_0^2+\epsilon_2^2)+2\kappa_1(\epsilon_0\Q_0
+\epsilon_2\Q_2)]+[\epsilon_0^2\epsilon_2^2
+2\kappa_1\epsilon_0\epsilon_2(\epsilon_0\Q_2+\epsilon_2\Q_0)]=0.$$
Using here the expressions (\ref{sums}) for $\Q_0,\,\Q_2$ and the
self-consistent value of the strength constant (A.3), we find
$$E^4-E^2(1-\alpha/2)(\epsilon_0^2+\epsilon_2^2)
+(1-\alpha)\epsilon_0^2\epsilon_2^2=0,$$ or
\begin{equation}
\label{Secpol}
\Omega^4-\Omega^2(2-\alpha)\omega_+^2+(1-\alpha)\omega_-^4=0,
\end{equation}
with the notation $\omega_+^2=\omega_x^2+\omega_z^2$ and
$\omega_-^4=(\omega_x^2-\omega_z^2)^2$.
This result coincides with that of \cite{Hamam}. By a trivial
rearrangement of the terms in (\ref{Secpol}) one obtains the useful
relation
\begin{equation}
\label{Relat}
\Omega^2(\Omega^2-\omega_+^2)=
(1-\alpha)(\Omega^2\omega_+^2-\omega_-^4).
\end{equation}
Inserting expressions (A.3) for $\omega_x^2,\,\omega_z^2$
into (\ref{Secpol}), we find
$\omega_+^2=2\bar\omega^2(1+\delta/3),\,
\omega_-^4=4\delta^2\bar\omega^4$
and reproduce formula (\ref{harac2}) for the isovector case
$$\Omega^4-2\Omega^2\bar{\omega}^2(2-\alpha)(1+\delta/3)
+4\bar{\omega}^4(1-\alpha)\delta^2=0.$$
Taking here $\alpha=1$ we reproduce formula (\ref{haracis2}) for
the isoscalar case
$$\Omega^4-2\Omega^2\bar{\omega}^2(1+\delta/3)=0.$$

\subsection{B(E2)-factors}

 According to \cite{Ring}, the transition probability for the one-body
operator $\di{\hat F=\sum_{s=1}^A\hat f_s}$
is calculated by means of the formulae
\begin{equation}
\label{matelem}
<0|\hat F^{\tau}|\nu>=\sum_{mi}(f_{im}^{\tau}X_{mi}^{\tau,\nu}
+f_{mi}^{\tau}Y_{mi}^{\tau,\nu}),\quad
<\nu|\hat F^{\tau}|0>=\sum_{mi}(f_{mi}^{\tau}X_{mi}^{\tau,\nu}
+f_{im}^{\tau}Y_{mi}^{\tau,\nu}).
\end{equation}
Quadrupole excitations are described by the operator (\ref{Oelec})
with $\hat f_{2\mu}=er^2Y_{2\mu}=\tilde e\Q$, where
$\tilde e=e\sqrt{\frac{5}{16\pi}}$.
The expressions for $X_{mi}^{\tau},\,Y_{mi}^{\tau}$ are given by
formulae (\ref{XY}). Combining these results we get
\begin{equation}
<0|\hat F_{21}^{\rm p}|\nu>=2\tilde e
K_{\nu}^{\rm p}\sum_{mi}|\Q_{mi}^{\rm p}|^2
\frac{\epsilon_{mi}^{\rm p}}{E_{\nu}^2-(\epsilon_{mi}^{\rm p})^2}=
2\tilde eK_{\nu}^{\rm p}S_{\nu}^{\rm p}=\tilde eC_{\nu}^{\rm p}.
\label{probab}
\end{equation}
The constant $C_{\nu}^{\rm p}$
is determined by the normalization condition
$$\delta_{\nu,\nu'}=\sum_{mi,\tau}(X_{mi}^{\tau,\nu*}X_{mi}^{\tau,\nu'}
-Y_{mi}^{\tau,\nu*}Y_{mi}^{\tau,\nu'}),$$
that gives
\begin{equation}
\frac{1}{(C_{\nu}^{\rm p})^2}=E_{\nu}\sum_{mi}\left[
\frac{|\Q_{mi}^{\rm p}|^2}{(S_{\nu}^{\rm p})^2}
\frac{\epsilon_{mi}^{\rm p}}{[E_{\nu}^2-(\epsilon_{mi}^{\rm p})^2]^2}
+\frac{(C_{\nu}^{\rm n})^2}{(C_{\nu}^{\rm p})^2}
\frac{|\Q_{mi}^{\rm n}|^2}{(S_{\nu}^{\rm n})^2}
\frac{\epsilon_{mi}^{\rm n}}{[E_{\nu}^2-(\epsilon_{mi}^{\rm n})^2]^2}
\right].
\label{norma}
\end{equation}
The ratio $C^{\rm n}/C^{\rm p}$ is determined by any of the equations
(\ref{Cnp}):
\begin{equation}
\frac{C^{\rm n}}{C^{\rm p}}
=\frac{1-2S^{\rm p}\kappa}{2S^{\rm p}\bar\kappa}
=\frac{2S^{\rm n}\bar\kappa}{1-2S^{\rm n}\kappa}.
\label{Cn/Cp}
\end{equation}
Formula (\ref{norma}) is considerably simplified by the approximation
(\ref{Apr4}), when $S^{\rm p}=S^{\rm n}\equiv S/2,\,
\epsilon_{mi}^{\rm p}=\epsilon_{mi}^{\rm n},\,
\Q_{mi}^{\rm p}=\Q_{mi}^{\rm n}$. Applying the second forms of
formulae (\ref{Secis},\,\ref{Seciv}) it is easy to find that in this
case $C^{\rm n}/C^{\rm p}=\pm 1$. As a result,
the final expression for $B(E2)$ value is
\begin{equation}
B(E2)_{\nu}=2|<0|\hat F_{21}^{\rm p}|\nu>|^2
=2\tilde e^2\left(16E_{\nu}\kappa_1^2\sum_{mi}|\Q_{mi}|^2
\frac{\epsilon_{mi}}{(E_{\nu}^2-\epsilon_{mi}^2)^2}\right)^{-1}.
\label{BE2}
\end{equation}
With the help of formulae (\ref{sums}) this expression can be
transformed into
\begin{eqnarray}
B(E2)_{\nu}&=&\frac{5}{8\pi}\frac{e^2Q_{00}}{m\bar\omega^2\alpha^2E_{\nu}}
\left[\frac{\epsilon_0^2}{(E_{\nu}^2-\epsilon_0^2)^2}
     +\frac{\epsilon_2^2}{(E_{\nu}^2-\epsilon_2^2)^2}\right]^{-1}
\nonumber\\
&=&\frac{5}{16\pi}\frac{e^2\hbar Q_{00}}{m\bar\omega^2\Omega_{\nu}}
\frac{(\Omega_{\nu}^2\omega_+^2 -\omega_-^4)^2}
{\Omega_{\nu}^4\omega_+^2 -2\Omega_{\nu}^2\omega_-^4
+\omega_+^2\omega_-^4}.
\label{BE2rpa}
\end{eqnarray}
At first sight, this expression has nothing in common with
(\ref{BE2gen}).
Nevertheless, it can be shown that they are identical. To this end,
we analyze carefully the denominator of the last expression in
(\ref{BE2rpa}). Summing it with the secular equation (\ref{Secpol})
(multiplied by $\omega_+^2$), which
obviously does not change its value, we find after elementary
combinations
\begin{eqnarray}
{\rm Denom}&=&\Omega_{\nu}^4\omega_+^2 -2\Omega_{\nu}^2\omega_-^4
+\omega_+^2\omega_-^4
+\omega_+^2[\Omega_{\nu}^4-\Omega_{\nu}^2(2-\alpha)\omega_+^2
+(1-\alpha)\omega_-^4]
\nonumber\\
&=&\omega_+^2\Omega_{\nu}^2[2\Omega_{\nu}^2-(2-\alpha)\omega_+^2]
-\omega_-^4[2\Omega_{\nu}^2-(2-\alpha)\omega_+^2]
\nonumber\\
&=&(\Omega_{\nu}^2\omega_+^2-\omega_-^4)
[2\Omega_{\nu}^2-(2-\alpha)\omega_+^2].
\label{denom}
\end{eqnarray}
This result allows us to write the final expression as
\begin{eqnarray}
B(E2)_{\nu}=
\frac{5}{16\pi}\frac{e^2\hbar}{m\bar\omega^2}Q_{00}
\frac{\Omega_{\nu}^2\omega_+^2 -\omega_-^4}
{\Omega_{\nu}[2\Omega_{\nu}^2-(2-\alpha)\omega_+^2]},
\label{BE2fin}
\end{eqnarray}
which coincides with (\ref{BE2gen}) (we recall that
$\omega_+^2=2\bar\omega^2(1+\delta/3),\,
\omega_-^4=4\delta^2\bar\omega^4$).
By simple transformations
this formula is reduced to the result of Hamamoto and Nazarewicz
\cite{Hamam} (taking into account, that they published it without
the constant factor
$\di{\frac{5}{32\pi}\frac{e^2\hbar}{m\omega_0}Q_{00}^0}$).

\subsection{B(M1)-factors}

In accordance with formulae (\ref{Omagn}), (\ref{matelem}),
(\ref{XY}) the magnetic transition matrix element is given by
\begin{equation}
<0|\hat F^{\rm p}_{11}|\nu>=K_{\nu}^{\rm p}\sum_{mi}\left[
\frac{(\hat f^{\rm p}_{11})_{im}\Q_{im}^{\rm p*}}
{E_{\nu}-\epsilon_{mi}^{\rm p}}-
\frac{(\hat f^{\rm p}_{11})_{mi}\Q_{mi}^{\rm p*}}
{E_{\nu}+\epsilon_{mi}^{\rm p}}
\right].
\label{Mprob}
\end{equation}
As it is shown in Appendix B, the matrix element
$(f^{\rm p}_{11})_{im}$ is proportional to $\Q_{im}^{\rm p}$ (formula
(B.16). So, expression (\ref{Mprob}) is reduced to
\begin{eqnarray}
<0|\hat F^{\rm p}_{11}|\nu>&=&-K_{\nu}^{\rm p}
\frac{\tilde e\hbar}{2c\sqrt5}
(\omega_x^2-\omega_z^2)^{\rm p}\sum_{mi}\left[
\frac{\Q^{\rm p}_{im}\Q_{im}^{\rm p*}}
{\epsilon_{im}^{\rm p}(E_{\nu}-\epsilon_{mi}^{\rm p})}-
\frac{\Q^{\rm p}_{mi}\Q_{mi}^{\rm p*}}
{\epsilon_{mi}^{\rm p}(E_{\nu}+\epsilon_{mi}^{\rm p})}\right]
\nonumber\\
&=&K_{\nu}^{\rm p}\frac{\tilde e\hbar}{c\sqrt5}
(\omega_x^2-\omega_z^2)^{\rm p}E_{\nu}\sum_{mi}
\frac{|\Q^{\rm p}_{mi}|^2}
{\epsilon_{mi}^{\rm p}[E_{\nu}^2-(\epsilon_{mi}^{\rm p})^2]}.
\label{Magnet}
\end{eqnarray}
With the help of approximation (\ref{Apr4}) and the expressions
(\ref{sums}) for $\Q_0,\,\Q_2$ we find
\begin{eqnarray}
<0|\hat F^{\rm p}_{11}|\nu>
&=&\frac{C_{\nu}^{\rm p}}{2S_{\nu}^{\rm p}}
\frac{\tilde e\hbar}{c\sqrt5}(\omega_x^2-\omega_z^2)
\frac{Q_{00}}{2m\bar\omega^2}
(\frac{E_{\nu}}{E_{\nu}^2-\epsilon_0^2}
+\frac{E_{\nu}}{E_{\nu}^2-\epsilon_2^2})
\nonumber\\
&=&-2\kappa_1C_{\nu}^{\rm p}\frac{\tilde e}{c\sqrt5}
(\omega_x^2-\omega_z^2)\frac{Q_{00}}{m\bar\omega^2}
\frac{\Omega_{\nu}(\Omega_{\nu}^2-\omega_+^2)}
{\alpha(\Omega_{\nu}^2\omega_+^2-\omega_-^4)}
\nonumber\\
&=&\frac{C_{\nu}^{\rm p}}{2}\frac{\tilde e}{c\sqrt5}
(\omega_x^2-\omega_z^2)\frac{1-\alpha}{\Omega_{\nu}}.
\label{Magnet1}
\end{eqnarray}
Relation (\ref{Relat}) and the self-consistent value of the
strength constant $\kappa_1=\alpha\kappa_0$ were used in the last
step. For the magnetic transition probability we have
\begin{eqnarray}
B(M1)_{\nu}=2|<0|\hat F^{\rm p}_{11}|\nu>|^2
=2\frac{(C_{\nu}^{\rm p})^2}{4}\frac{\tilde e^2}{5c^2}
\omega_-^4\frac{(1-\alpha)^2}{\Omega_{\nu}^2}
=\frac{\omega_-^4}{20c^2}
\frac{(1-\alpha)^2}{\Omega_{\nu}^2}B(E2)_{\nu}.
\label{BM1rpa}
\end{eqnarray}
This relation between $B(M1)$ and $B(E2)$ was also found (up to the
factor $1/(20c^2)$) by Hamamoto and Nazarewicz \cite{Hamam}.
Substituting expression (\ref{BE2fin}) for $B(E2)$ into (\ref{BM1rpa})
we reproduce (with the help of relation (\ref{Relat})) formula
(\ref{BM1gen}).

\subsection{``Synthetic" scissors and spurious state}

 The nature of collective excitations calculated with the method of
Wigner function moments is quite easily revealed analyzing the
roles of collective variables describing the
phenomenon. The solution of this problem in the RPA approach is not so
obvious. That is why the nature of the low-lying states has often been
established by considering overlaps of these states with the "pure
scissors state" \cite{Hilt86,Dieper} or "synthetic state" \cite{Hamam}
produced by the action of the scissors operator which antirotates
proton versus neutron distributions
$$\hat S_x=\N^{-1}(\langle{I_x^{\rm n}}^2\rangle\hat I_x^{\rm p}-
\langle{I_x^{\rm p}}^2\rangle\hat I_x^{\rm n})$$
on the ground state
$$|Syn>=\hat S_x|0>.$$
In the considered model the overlap of the ``synthetic" state with
the real scissors mode (and with IVGQR) can be calculated
analytically. Let us at first
modify the definition of the ``synthetic" state.
Due to axial symmetry one can use the $\hat I_y^{\tau}$ component
instead of $\hat I_x^{\tau}$, or any of their linear combinations,
for example, the $\mu=1$ component of the magnetic operator
$\hat F_{1\mu}^{\tau}$, which is much more convenient for us.
The terms $\langle{I_x^{\tau}}^2\rangle$ are introduced to
ensure the orthogonality of the synthetic scissors to the spurious
state $|Sp>=(\hat I^{\rm n}+\hat I^{\rm p})|0>$.
However, we do not need these
terms because the collective states $|\nu>$ of our model are already
orthogonal to $|Sp>$ (see below); hence, the overlaps $<Syn|\nu>$
will be free from any admixtures of $|Sp>$. So, we use the
following definitions of the synthetic and spurious states:
$$|Syn>=\N^{-1}(\hat F_{11}^{\rm p}-\hat F_{11}^{\rm n})|0>,\qquad
|Sp>=(\hat F_{11}^{\rm p}+\hat F_{11}^{\rm n})|0>.$$

 Let us demonstrate the orthogonality
of the spurious state to all the rest of the states $|\nu>$.
As the first step it is necessary to show that the secular equation
(\ref{Secul1}) has the solution $E=0$.
We need the expression for $S^{\tau}(E=0)\equiv S^{\tau}(0)$.
In accordance with (\ref{S}), we have
$$S^{\tau}(E)=\left[\frac{\epsilon_0\Q_0}{E^2-\epsilon_0^2}+
\frac{\epsilon_2\Q_2}{E^2-\epsilon_2^2}\right]^{\tau},\quad
S^{\tau}(0)=-\left[\frac{\Q_0}{\epsilon_0}+
\frac{\Q_2}{\epsilon_2}\right]^{\tau}.$$
 The expressions for $\Q_0^{\tau},\,\Q_2^{\tau}$ are easily extracted
from formulae (B.10), (B.11):
\begin{equation}
\Q_0^{\tau}=\frac{\hbar}{m}Q_{00}^{\tau}\left[
\frac{1+\frac{4}{3}\delta}{\omega_x}
-\frac{1-\frac{2}{3}\delta}{\omega_z}\right]^{\tau},
\quad \Q_2^{\tau}=\frac{\hbar}{m}Q_{00}^{\tau}\left[
\frac{1+\frac{4}{3}\delta}{\omega_x}
+\frac{1-\frac{2}{3}\delta}{\omega_z}\right]^{\tau}.
\label{D0D2}
\end{equation}
So we find
\begin{eqnarray}
 S^{\tau}(0)&=&
-\frac{\hbar}{m}Q_{00}^{\tau}\left[
\frac{1+\frac{4}{3}\delta}{\omega_x}
(\frac{1}{\epsilon_2}+\frac{1}{\epsilon_0})
+\frac{1-\frac{2}{3}\delta}{\omega_z}
(\frac{1}{\epsilon_2}-\frac{1}{\epsilon_0})
\right]^{\tau}
\nonumber\\
&=&-\frac{\hbar^2}{m}
\frac{4\delta^{\tau}Q_{00}^{\tau}}
{\epsilon_2^{\tau}\epsilon_0^{\tau}}
=-\frac{1}{m}
\frac{3Q_{20}^{\tau}}
{(\omega_x^2-\omega_z^2)^{\tau}},
\label{S0}
\end{eqnarray}
where, in accordance with (B.12),
\begin{equation}
(\omega_x^2-\omega_z^2)^{\rm p}=
-\frac{6}{m}(\kappa Q_{20}^{\rm p}+\bar\kappa Q_{20}^{\rm n}),\quad
(\omega_x^2-\omega_z^2)^{\rm n}=
-\frac{6}{m}(\kappa Q_{20}^{\rm n}+\bar\kappa Q_{20}^{\rm p}).
\label{omeg-}
\end{equation}
Finally, we get
$$2S^{\rm p}(0)=\frac{Q_{20}^{\rm p}}{\kappa Q_{20}^{\rm p}
+\bar\kappa Q_{20}^{\rm n}},\quad
1-2S^{\rm p}(0)\kappa=\frac{\bar\kappa Q_{20}^{\rm n}}
{\kappa Q_{20}^{\rm p}+\bar\kappa Q_{20}^{\rm n}},$$
$$2S^{\rm n}(0)=\frac{Q_{20}^{\rm n}}{\kappa Q_{20}^{\rm n}
+\bar\kappa Q_{20}^{\rm p}},\quad
1-2S^{\rm n}(0)\kappa=\frac{\bar\kappa Q_{20}^{\rm p}}
{\kappa Q_{20}^{\rm n}+\bar\kappa Q_{20}^{\rm p}}.$$
It is easy to see that substituting these expressions into
(\ref{Secul1}) we obtain an identity; therefore, the secular
equation has a zero energy solution.

For the second step it is necessary to calculate the overlap
$<Sp|\nu>$.
 Summing (\ref{Magnet}) with an
analogous expression for neutrons, we get
\begin{eqnarray}
<Sp|\nu>
&=&\frac{\tilde e\hbar}{c\sqrt5}E_{\nu}\sum_{\tau}
K_{\nu}^{\tau}
(\omega_x^2-\omega_z^2)^{\tau}
\sum_{mi}\frac{|\Q_{mi}^{\tau}|^2}
{\epsilon_{mi}^{\tau}(E_{\nu}^2-\epsilon_{mi}^2)^{\tau}}
\nonumber\\
&=&\frac{\tilde e\hbar}{c\sqrt5}E_{\nu}\sum_{\tau}
K_{\nu}^{\tau}
(\omega_x^2-\omega_z^2)^{\tau}
\sum_{mi}\frac{|\Q_{mi}^{\tau}|^2\epsilon_{mi}^{\tau}}
{(\epsilon_{mi}^2)^{\tau}(E_{\nu}^2-\epsilon_{mi}^2)^{\tau}}.
\label{ortog}
\end{eqnarray}
Applying the algebraical identity
$$\frac{1}{\epsilon^2(E^2-\epsilon^2)}=
\frac{1}{E^2}(\frac{1}{\epsilon^2}+
\frac{1}{E^2-\epsilon^2})$$
and remembering the definition (\ref{S}) of $S^{\tau}$ we
rewrite (\ref{ortog}) as
\begin{eqnarray}
<Sp|\nu>
&=&\frac{\tilde e\hbar}{c\sqrt5E_{\nu}}\sum_{\tau}
K_{\nu}^{\tau}
(\omega_x^2-\omega_z^2)^{\tau}(S^{\tau}-S^{\tau}(0))
\nonumber\\
&=&\frac{\tilde e\hbar}{c\sqrt5}\frac{K_{\nu}^{\rm p}}{E_{\nu}}
\left[
(\omega_x^2-\omega_z^2)^{\rm p}
(S^{\rm p}-S^{\rm p}(0))
+(\omega_x^2-\omega_z^2)^{\rm n}
(S^{\rm n}-S^{\rm n}(0))
\frac{K_{\nu}^{\rm n}}{K_{\nu}^{\rm p}}
\right].
\label{ortogo}
\end{eqnarray}
In accordance with (\ref{CSK}) and (\ref{Cn/Cp}),
\begin{equation}
\frac{K_{\nu}^{\rm n}}{K_{\nu}^{\rm p}}=
\frac{1-2S^{\rm p}\kappa}{2S^{\rm n}\bar\kappa}.
\label{Kn/Kp}
\end{equation}
Noting now (see formula (\ref{S0})) that
$ (\omega_x^2-\omega_z^2)^{\tau}S^{\tau}(0)
=-\frac{3}{m}Q_{20}^{\tau}$ and
taking into account relations (\ref{omeg-}), we find
\begin{eqnarray}
<Sp|\nu>
&=&\beta
\left\{
[(\kappa Q_{2}^{\rm p}+\bar\kappa Q_{2}^{\rm n})2S^{\rm p}
-Q_{2}^{\rm p}]
+[(\kappa Q_{2}^{\rm n}+\bar\kappa Q_{2}^{\rm p})2S^{\rm n}
-Q_{2}^{\rm n}]
\frac{1-2S^{\rm p}\kappa}{2S^{\rm n}\bar\kappa}
\right\}
\nonumber\\
&=&\beta
\left\{
[(2S^{\rm p}\kappa-1)Q_{2}^{\rm p}
+2S^{\rm p}\bar\kappa Q_{2}^{\rm n}]
+[(2S^{\rm n}\kappa -1)Q_{2}^{\rm n}
+2S^{\rm n}\bar\kappa Q_{2}^{\rm p})]
\frac{1-2S^{\rm p}\kappa}{2S^{\rm n}\bar\kappa}
\right\}
\nonumber\\
&=&\beta
\left\{
2S^{\rm p}\bar\kappa Q_{2}^{\rm n}
+(2S^{\rm n}\kappa -1)Q_{2}^{\rm n}
\frac{1-2S^{\rm p}\kappa}
{2S^{\rm n}\bar\kappa}
\right\}
\nonumber\\
&=&\beta \frac{Q_{2}^{\rm n}}{2S^{\rm n}\bar\kappa}
\left\{
2S^{\rm n}\bar\kappa
2S^{\rm p}\bar\kappa
-(1-2S^{\rm n}\kappa)
(1-2S^{\rm p}\kappa)
\right\}=0,
\label{ortogon}
\end{eqnarray}
where $\di\beta=-\frac{3}{m}
\frac{\tilde e\hbar}{c\sqrt5}\frac{K_{\nu}^{\rm p}}{E_{\nu}}$
and $Q_2\equiv Q_{20}$.
The expression in the last curly brackets coincides obviously with the
secular equation (\ref{Secul1}) that proves the orthogonality of the
spurious state to all physical states of the considered model. So we
can conclude that strictly speaking this is not a spurious state,
but one of the exact eigenstates of the model corresponding to the
integral of motion $I^{\rm n}+I^{\rm p}$.
The same conclusion was made
by N. Lo Iudice \cite{Lo96} who solved this problem approximately
with the help of several assumptions (the small deformation limit, for
 example).

The problem of the "spurious" state being solved, the calculation of
overlaps $<Syn|\nu>$ becomes trivial. Really, we have shown that
$<0|\hat F^{\rm n}_{11}+\hat F^{\rm p}_{11}|\nu>=0$, hence
$<0|\hat F^{\rm n}_{11}|\nu>=\\
-<0|\hat F^{\rm p}_{11}|\nu>$. Then
$<Syn|\nu>=\N^{-1}<0|\hat F^{\rm p}_{11}-\hat F^{\rm n}_{11}|\nu>=
2\N^{-1}<0|\hat F^{\rm p}_{11}|\nu>$ and
\begin{equation}
U^2\equiv|<Syn|\nu>|^2=2\N^{-2}B(M1)_{\nu}.
\label{U2}
\end{equation}
The nontrivial part of the problem is the calculation of the
normalization factor $\N$. It is important not to forget about the
time dependence of the synthetic state which should be determined
by the external field:
$$|Syn(t)>=\N^{-1}[
(\hat F_{11}^{\rm p}-\hat F_{11}^{\rm n})e^{-i\Omega t}
+(\hat F_{11}^{\rm p}-\hat F_{11}^{\rm n})^{\dagger}e^{i\Omega t}
]|0>.$$
 Then we have
\begin{eqnarray}
\N^2&=&2<0|(\hat F^{\rm p}_{11}-\hat F^{\rm n}_{11})^{\dagger}
(\hat F^{\rm p}_{11}-\hat F^{\rm n}_{11})|0>
\nonumber\\
&=&
2\sum_{ph}<0|(\hat F^{\rm p}_{11}-\hat F^{\rm n}_{11})^{\dagger}|ph>
<ph|(\hat F^{\rm p}_{11}-\hat F^{\rm n}_{11})|0>=
2\sum_{ph}|<ph|(\hat F^{\rm p}_{11}-\hat F^{\rm n}_{11})|0>|^2
\nonumber\\
&=&
2\sum_{\tau,ph}|<ph|\hat F^{\tau}_{11}|0>|^2=
2\sum_{\tau,ph}|(f^{\tau}_{11})_{ph}|^2.
\label{Norm}
\end{eqnarray}
With the help of relation (B.16) we find
\begin{eqnarray}
\N^2
&=&\frac{2}{5}
(\frac{e\hbar}{2c})^2 \sum_{\tau,ph} \left(
\omega_-^4
\frac{|<ph|r^2Y_{21}|0>|^2}{\epsilon^2_{ph}}\right)^{\tau}
\nonumber\\
&=&\frac{1}{8\pi}
(\frac{e\hbar}{2c})^2
 \sum_{\tau}
(\omega_-^4)^{\tau}
\left(
\frac{\Q_0}{\epsilon^2_0}
+\frac{\Q_2}{\epsilon^2_2}
\right)^{\tau}.
\label{Norma}
\end{eqnarray}
 Expressions for $\Q_0^{\tau},\,\Q_2^{\tau},\,
\omega_x^{\tau},\,\omega_z^{\tau}$ are given by formulae
(\ref{D0D2}), (B.12). To get a definite number, it is necessary
to make some assumption concerning the relation between neutron and
proton equilibrium characteristics. As usual, we apply the
approximation (\ref{Apr4}), i.e., suppose
$Q_{00}^{\rm n}=Q_{00}^{\rm p},\,Q_{20}^{\rm n}=Q_{20}^{\rm p}$.
It is easy to check that in this case formulae for
$\omega_{x,z}^{\tau}$ are reduced to the ones for the isoscalar case,
namely (A.3), and $\Q_0^{\tau}=\Q_0/2,\,\Q_2^{\tau}=\Q_2/2$, where
$\Q_0$ and $\Q_2$ are given by (\ref{sums}). So we get
\begin{eqnarray}
\N^2
=\frac{\omega_-^4}{8\pi}
(\frac{e\hbar}{2c})^2
\frac{Q_{00}}{m\bar\omega^2}
\left(\frac{1}{\epsilon_0}
+\frac{1}{\epsilon_2}\right)
=\frac{\delta}{2\pi}
\frac{m\omega_x}{\hbar}Q_{00}\mu_N^2.
\label{Norma2}
\end{eqnarray}
The estimation of the overlap for $^{156}$Gd with $\delta=0.27$ gives
$\N^2=34.72\mu_N^2$ and $U^2=0.53$ (see eq. (\ref{U2})), that is two
times larger than the
result of \cite{Hilt86} obtained in QRPA calculations with the Skyrme
forces. The disagreement can naturally be attributed to the difference
in forces and especially to the lack of pair correlations in our
approach.

In the small deformation
 limit $U^2=\frac{1}{2}\sqrt{\frac{3}{2}}\approx 0.6.$
 This is the maximum possible overlap of the "pure" (or "synthetic")
scissors with the real scissors. Increasing $\delta$ and /or
taking into account pairing correlations decreases its value, that is
confirmed by numerous microscopic calculations with various forces
\cite{Zaw}. Such small overlap leads inevitably to the conclusion,
that the original model of counter rotating rigid rotors \cite{Lo2000}
has not very much in common with the real scissors mode, the correct
description of which requires the proper treatment of the Fermi surface
deformation and the coupling with IVGQR.

\subsection{Superdeformation}

A certain drawback of our approach is that, so
far, we have not included the superfluidity into our description.
On the other hand, our formulae (\ref{Omeg2fin}, \ref{BM1gen}) can be
successfully used
for the description of superdeformed nuclei where the pairing
is very weak \cite{Hamam,Lo2000}. For example, applying them to the
superdeformed nucleus $^{152}$Dy ($\delta\simeq 0.6,\,
\hbar\omega_0=41/A^{1/3} \rm{MeV}$), we get
$$E_{iv}=20.8\, {\rm MeV},\quad\quad B(M1)_{iv}=15.9\, \mu_N^2$$
for the isovector GQR and
$$E_{sc}=4.7\, {\rm MeV},\quad\quad B(M1)_{sc}=20.0\, \mu_N^2$$
for the scissors mode. There are not many results of other
calculations to compare with. As a matter of fact, there are only two
papers considering this problem.

The phenomenological TRM model \cite{Lo2000} predicts
$$E_{iv}\simeq26\, {\rm MeV},\quad B(M1)_{iv}\simeq26\, \mu_N^2,\quad
E_{sc}\simeq6.1\, {\rm MeV},\quad B(M1)_{sc}\simeq22\, \mu_N^2.$$
The only existing microscopic calculation \cite{Hamam}
in the framework of QRPA with separable forces gives
$$E_{iv}\simeq28\, {\rm MeV},\quad B(M1)_{iv}\simeq37\, \mu_N^2,\quad
E_{sc}\simeq5-6\, {\rm MeV},\quad B(M1)_{1^+}\simeq23\, \mu_N^2.$$
Here $B(M1)_{1^+}$ denotes
the total $M1$ orbital strength carried by the
calculated $K^{\pi}=1^+$ QRPA excitations modes in the energy region
below 20 MeV.

It is easy to see that in the case of IVGQR one can speak, at least,
about qualitative agreement. Our
results for $E_{sc}$ and $B(M1)_{sc}$ are in good agreement with
that of phenomenological model and with $E_{sc}$ and $B(M1)_{1^+}$ of
Hamamoto and Nazarewicz.

 It is possible to extract from the histogram
of \cite{Hamam} the value of the overlap of calculated low-lying
$1^+$ excitations with the synthetic scissors state:
$|<Syn|1^+>|^2 \approx 0.4.$ The result of our calculation
$U^2=0.43$ agrees with it very well.
 So, the comparison of our calculations with that of QRPA shows, that
we have excellent agreement in superdeformed nuclei and rather large
disagreement in moderately deformed nuclei. On the other hand it is
known \cite{Hamam}, that
pairing is very weak at the superdeformation and becomes
important at moderate deformations. Therefore, as a consequence
the correct treatment of pair correlations is
important for an accurate description of the scissors mode. This
shall be the subject of future work.

\section{WFM versus RPA}

In this section we want to establish the precise relation between the
WFM and RPA methods. Though it can be guessed that working in both
cases with the full configuration space, i.e. with a complete set of
particle and hole states in the case of RPA and with all moments of
all ranks in the case of WFM, identical results shall be obtained, it
is still important to study the detailes of this relation because the
behaviour of WFM and RPA under truncation of the space turns out to be
radically different.
The exact relations between the RPA amplitudes $X_{ph}$ and $Y_{ph}$
and the respective WFM variables can be established with the help of
the linear response theory.

 Let us consider the response of the system to a
weak external time-dependent field
\begin{equation}
\label{Exfield}
\hat W(t)=\hat W\,{\rm exp}(-i\Omega t)
+\hat W^{\dagger}\,{\rm exp}(i\Omega t)
\end{equation}
with $\hat W=\sum_{kq}w_{kq}a^{\dagger}_k a_q$. The linear response
is then given by \cite{Ring}:
$$\rho^{(1)}_{kq}(t)=
\sum_{k'q'}\left[R_{kq,k'q'}(\Omega)e^{-i\Omega t}
+R_{qk,k'q'}^*(\Omega)e^{i\Omega t}\right]w_{k'q'},$$
where
$$R_{kq,k'q'}(\Omega)=\sum_{\nu}\left(
\frac{<0|a^{\dagger}_q a_k|\nu><\nu|a^{\dagger}_{k'} a_{q'}|0>}
{\hbar(\Omega-\Omega_{\nu})}-
\frac{<0|a^{\dagger}_{k'} a_{q'}|\nu><\nu|a^{\dagger}_{q} a_{k}|0>}
{\hbar(\Omega+\Omega_{\nu})}\right)$$
is the RPA response function \cite{Ring} and the index pairs $kq$
and $k'q'$ are restricted to particle hole and hole particle pairs.
Indices $p,q$ include spin and isospin quantum numbers $\sigma$ and
$\tau$.
For the change of the average value of an arbitrary operator we have:
\begin{equation}
\label{varF}
\delta<\Psi|\hat F|\Psi>=\sum_{kq}f_{qk}\rho_{kq}^{(1)}.
\end{equation}

We now are ready to analyze the WFM variables. The first one is
$$ R_{\lambda\mu}^{\tau}(t)=
2(2\pi\hbar)^{-3}\int\! d^3p\,\int\! d^3r
r_{\lambda\mu}^{2}f^{\tau}(\br,\bp,t).$$
Using here definitions of the Wigner function, the $\delta$-function
and the density matrix \cite{Ring}
\begin{equation}
\label{densmat}
\rho(\br\sigma\tau,\br'\sigma'\tau')
=\sum_{kq}\phi_k^*(\br'\sigma'\tau')\phi_q(\br\sigma\tau)
<\Psi|a_k^{\dagger}a_q|\Psi>
\end{equation}
we find
\begin{eqnarray}
 R_{\lambda\mu}^{\tau}(t)&=&
\frac{2}{(2\pi\hbar)^{3}}
\int\! d^3r\,
r_{\lambda\mu}^{2} \int d^3s\int\! d^3p
\: \exp(- i\bp\cdot
 \bs/\hbar)\rho^{\tau}(\br+\frac{\bs}{2}, \br-\frac{\bs}{2},t)
\nonumber\\
&=&2 \int\! d^3r\, r_{\lambda\mu}^{2}\rho^{\tau}(\br, \br,t)
=\sum_{\sigma} \int\! d^3r\, r_{\lambda\mu}^{2}
\rho(\br\sigma\tau,\br\sigma\tau,t)
\nonumber\\
&=&\sum_{kq}\sum_{\sigma} \int\! d^3r\, r_{\lambda\mu}^{2}
\phi_k^*(\br\sigma\tau)\phi_q(\br\sigma\tau)
<\Psi|a_k^{\dagger}a_q|\Psi>
\nonumber\\
&=&\sum_{kq}(r_{\lambda\mu}^{2})_{kq}^{\tau}
<\Psi|a_k^{\dagger}a_q|\Psi>
=<\Psi|\sum_{kq}(r_{\lambda\mu}^{2})_{kq}^{\tau}a_k^{\dagger}a_q|\Psi>
\nonumber\\
&=&<\Psi|\sum_{s=1}^{N_{\tau}} r_{\lambda\mu}^{2}(s)|\Psi>
=<\Psi|\hat R_{\lambda\mu}^{\tau}|\Psi>,
\label{R2rho}
\end{eqnarray}
i.e. this is just the ground state expectation value of the operator
$\di\hat R_{\lambda\mu}=\sum_{s=1}^A(r^2_s)_{\lambda\mu}$.
In accordance with (\ref{varF}) the variation of this variable is
\begin{eqnarray}
 \delta R_{\lambda\mu}^{\tau}(t)&=&
 \R_{\lambda\mu}^{\tau}(t)
=\sum_{kq}(r_{\lambda\mu}^{2})_{kq}^{\tau}\rho^{(1)}_{qk}(t)
\nonumber\\
&=&\sum_{\nu}(<0|\hat R_{\lambda\mu}^{\tau}|\nu>c_{\nu}
-<\nu|\hat R_{\lambda\mu}^{\tau}|0>\bar c_{\nu})e^{-i\Omega t}
\nonumber\\
&&+\sum_{\nu}(<\nu|\hat R_{\lambda\mu}^{\tau}|0>c_{\nu}^*
-<0|\hat R_{\lambda\mu}^{\tau}|\nu>\bar c_{\nu}^*)e^{i\Omega t},
\label{deltaR}
\end{eqnarray}
where
\begin{equation}
\label{Cnu}
 c_{\nu}=\frac{<\nu|\hat W|0>}{\hbar(\Omega-\Omega_{\nu})}
=\sum_{kq}
\frac{<\nu|a^{\dagger}_k a_q|0>}{\hbar(\Omega-\Omega_{\nu})}
w_{kq},\quad
\bar c_{\nu}=\frac{<0|\hat W|\nu>}{\hbar(\Omega+\Omega_{\nu})}
=\sum_{kq}
\frac{<0|a^{\dagger}_k a_q|\nu>}{\hbar(\Omega+\Omega_{\nu})}
w_{kq}.
\end{equation}

Equation (\ref{deltaR}) demonstrates in
an obvious way the structure of the variables
$\delta R_{\lambda\mu}$. They are
linear combinations of the
transition matrix elements $<0|\hat R_{\lambda\mu}|\nu>$
which are, in turn, linear combinations of the RPA
amplitudes $X_{kq}, Y_{kq}$.
Introducing the notation
$\di \hat L_{\lambda\mu}=\sum_{s=1}^A(r_s\hat p_s)_{\lambda\mu}$
and
$\di \hat P_{\lambda\mu}=\sum_{s=1}^A(\hat p_s^2)_{\lambda\mu}$
we can proceed in a similar way with
$$ L_{\lambda\mu}^{\tau}(t)=
2(2\pi\hbar)^{-3}\int\! d^3p\,\int\! d^3r
(rp)_{\lambda\mu}f^{\tau}(\br,\bp,t)$$
and
$$ P_{\lambda\mu}^{\tau}(t)=
2(2\pi\hbar)^{-3}\int\! d^3p\,\int\! d^3r
p_{\lambda\mu}^{2}f^{\tau}(\br,\bp,t).$$

Inserting the expressions for $\delta R_{\lambda\mu},
\delta L_{\lambda\mu}$ into the first equation of the
set (\ref{quadr}) we find
$$-i\Omega\sum_{\nu}(<0|\hat R_{\lambda\mu}|\nu>c_{\nu}
-<\nu|\hat R_{\lambda\mu}|0>\bar c_{\nu})
=\frac{2}{m}\sum_{\nu}(<0|\hat L_{\lambda\mu}|\nu>c_{\nu}
-<\nu|\hat L_{\lambda\mu}|0>\bar c_{\nu}).$$
It is sufficient to consider only the part with the $e^{-i\Omega t}$
time dependence. Multiplying
this equation by $(\Omega-\Omega_{\nu})$ and taking the
limit $\Omega\to\Omega_{\nu}$ we find the equation
\begin{equation}
\label{TME}
-i\Omega_{\nu}<0|\hat R_{\lambda\mu}|\nu>
=\frac{2}{m}<0|\hat L_{\lambda\mu}|\nu>,
\end{equation}
which can be called as the dynamical equation for the transition
matrix element
$<0|\hat R_{\lambda\mu}|\nu>$.
 Naturally, in the same
way the dynamical equations for transition matrix elements
$<0|\hat L_{\lambda\mu}|\nu>$ and
$<0|\hat P_{\lambda\mu}|\nu>$ can be extracted from the second and
third (linearized) equations of (\ref{quadr}).

Now we can show, that exactly the same dynamical equations for
Transition Matrix Elements (TME) can be derived from RPA equations.
To this end we combine the RPA equations (\ref{DD})
 with the definition (\ref{matelem}) of matrix elements:
\begin{equation}
\label{Equmot}
\hbar\Omega_{\nu}\sum_{mi}(f_{im}^{\tau}X_{mi}^{\tau,{\nu}}
+f_{mi}^{\tau}Y_{mi}^{\tau,{\nu}})=
\sum_{mi}\epsilon_{mi}(f_{im}^{\tau}X_{mi}^{\tau,{\nu}}
-f_{mi}^{\tau}Y_{mi}^{\tau,{\nu}})+
K^{\tau}_{\nu}\sum_{mi}(f_{im}^{\tau}\Q_{im}^{\tau*}
-f_{mi}^{\tau}\Q_{mi}^{\tau*}).
\end{equation}
 Taking into account the relations
$$\epsilon_{mi}f_{im}=[\hat f,H_0]_{im},\quad
\epsilon_{mi}f_{mi}=-[\hat f,H_0]_{mi},\quad$$
one rewrites this equation as
\begin{equation}
\label{Equmot2}
\hbar\Omega_{\nu}<0|\hat F^{\tau}|\nu>=
\sum_{mi}\{[\hat f^{\tau},H_0^{\tau}]_{im}X_{mi}^{\tau,{\nu}}
+[\hat f^{\tau},H_0^{\tau}]_{mi}Y_{mi}^{\tau,{\nu}}+
K^{\tau}_{\nu}(f_{im}^{\tau}\Q_{im}^{\tau*}
-f_{mi}^{\tau}\Q_{mi}^{\tau*})\}.
\end{equation}
The Hamiltonian of the axially deformed harmonic oscillator
corresponding to the mean field (\ref{potenirr}) is
\begin{equation}
\label{H0}
H_0^{\tau}(\br)=\sum_{s=1}^{N_{\tau}}\{\frac{\hat\bp_s^2}{2m}+
\frac{1}{2}m\,\omega^2\br_s^2+
Z_{20}^{\tau}(eq)r^2_{20}(s)\}.
\end{equation}

Let us consider the operator
$\hat f=\sqrt6\,r^2_{21}
=q_{21}=\Q.$
Calculating the commutator
$$[r^2_{21},H_0]=i\hbar\frac{2}{m}(r\hat p)_{21}$$
we find from (\ref{Equmot2}) the following equation
\begin{eqnarray}
\hbar\Omega_{\nu}<0|\sum_{s=1}^{N_{\tau}}(r^2_{21})_s^{\tau}|\nu>&=&
i\hbar\frac{2}{m}
\sum_{mi}
\{((r\hat p)_{21})_{im}^{\tau}X_{mi}^{\tau,{\nu}}+
((r\hat p)_{21})_{mi}^{\tau}Y_{mi}^{\tau,{\nu}}\}
\nonumber\\
&&+K^{\tau}_{\nu}\sum_{mi}
(\Q_{im}^{\tau}\Q_{im}^{\tau*}
-\Q_{mi}^{\tau}\Q_{mi}^{\tau*}).
\label{EquD}
\end{eqnarray}
Taking into account relations $(\Q^*)_{im}=(\Q)_{mi}^*$
and $|\Q_{mi}|^2=|\Q_{im}|^2$ we find, that the last sum in (\ref{EquD})
is equal to zero. Applying again formula (\ref{matelem})
we write (\ref{EquD}) as
\begin{equation}
-i\Omega_{\nu}<0|\hat R_{21}^{\tau}|\nu>
=\frac{2}{m}<0|\hat L_{21}^{\tau}|\nu>
\label{Dyneq}
\end{equation}
reproducing equation (\ref{TME}).
The dynamical equations for $<0|\hat L_{21}^{\tau}|\nu>$ and
$<0|\hat P_{21}^{\tau}|\nu>$ are obtained in a similar way by
considering the operators $\hat f=(r\hat p)^2_{21}$ and
$\hat f=\hat p^2_{21}$ respectively. As it could be expected, the
resulting equations reproduce the corresponding TME equations derived
from (\ref{quadr}) with the help of the limit procedure.

So, there exists
one-to-one correspondence between the set of dynamical equations for
WFM variables and the set of dynamical equations for
Transition Matrix Elements (TME). This correspondence makes obvious
the fact that both sets have the same eigenvalues. On the other hand
the TME equations are just linear combinations
of the RPA equations. Therefore we can conclude that RPA and WFM
approaches generate identical eigenvalues. In this sense both
approaches are equivalent in all aspects. This concerns for instance
also the transition probabilities. However, for this equivalence to
be exact, one needs to work in the full space in both approaches,
that is in the complete particle hole space in RPA and taking all
phase space moments of all powers in WFM, a task which can hardly be
tackled in general.

The difference of the two approaches then shows up if truncations of
the dimension of the equations have to be
operated. In RPA one usually solves the equations with a restricted
number of discrete particle hole pairs, i.e. the dimension of the RPA
matrix is finite (in some works the RPA equations for finite nuclei
are, however, solved in full space, including continuum states
\cite{Shlomo,Giai}).
The result of such a diagonalisation usually yields a huge number of
discrete eigenvalues approximating more or less the spectrum one
would obtain from a solution in the full space. For instance resonances
in the continuum (e.g. giant resonances) will be mocked up by a bunch
of discrete states whose envelope may simulate the full solution.
Reducing the dimension of the particle hole space too much may lead to
a situation where the full solution is only approximated rather badly
and in an uncontrolled manner.

In the WFM method the dynamical equations for Cartesian tensors of
the rank $n=2$ are coupled (by the interaction terms in
(\ref{sincyclic})) with dynamical equations for tensors of the rank
$n=3$, these equations being coupled with the ones for tensors of the
rank $n=4$ and so on up
to $n=\infty$. Here one hopes that the essential part of physics
is described by a small number of the lowest ranks tensors.
The hope is based on the assumption that the higher rank
tensors (moments) are responsible for the more refined details and
that neglecting them does not appreciably influence the
description of the more global physics which is described with
the lower ranks tensors.
This assumption is substantiated in past applications of the WFM
method to realistic situations with Skyrme forces for the description
of collective nuclear modes \cite{Bal,Piper,Duran}. In those works it
has indeed
been demonstrated that even with a very limited number of low rank
phase space moments one can faithfully reproduce the centroid position
of the collective states. In this sense the WFM method is rather
similar to the sum rule approach which works, however,
only in the cases when practically all strength is exhausted by
one state, whereas WFM method works also in situations when
the strength is distributed among a few excitations.
From these studies it is then permitted to assume that the
inclusion of higher and higher rank moments will just give raise to a
refinement of the gross structure obtained with the low rank tensors.
A formal convergence study of this type has been performed in
the infinite matter case \cite{Providen}
where it was indeed shown that the moment
method allows to approach the full solution in an optimized way.

The net result is, that WFM and RPA approximate the exact infinite
spectrum
into a finite number of eigenfrequencies with, however, different
convergence.

An analogous situation occurs with transition probabilities. Let us
analyze, for example, the expression (\ref{deltaR}) for the WFM
 variable
$\delta R_{\lambda\mu}^{\tau}(t)= \R_{\lambda\mu}^{\tau}(t)$.
Using the definition (\ref{Cnu}) of $c_{\nu}$ with the external field
operator $\hat W=\hat R^{\tau\dagger}_{\lambda\mu}$, we find
\begin{eqnarray}
\delta R_{\lambda\mu}^{\tau}(t)&=&
\sum_{\nu=1}^{N_c}\left(\frac{<0|\hat R_{\lambda\mu}^{\tau}|\nu>
<\nu|\hat R_{\lambda\mu}^{\tau \dagger}|0>}
{\hbar(\Omega-\Omega_{\nu})}-\frac{<\nu|\hat R_{\lambda\mu}^{\tau}|0>
<0|\hat R_{\lambda\mu}^{\tau \dagger}|\nu>}
{\hbar(\Omega+\Omega_{\nu})}\right)e^{-i\Omega t}
\nonumber\\
&=&\sum_{\nu=1}^{N_c}\left(
\frac{|<0|\hat R_{\lambda\mu}^{\tau}|\nu>|^2}
{\hbar(\Omega-\Omega_{\nu})}
-\frac{|<0|\hat R_{\lambda\mu}^{\tau \dagger}|\nu>|^2}
{\hbar(\Omega+\Omega_{\nu})}
\right)e^{-i\Omega t}.
\label{strength}
\end{eqnarray}
The summation limit $N_c$ depends on the
method of calculation. In the case of the exact solution
$N_c=\infty$, for RPA $N_c$ is usually of the order of several
$hundreds or thousands$, for WFM $N_c$ usually is not more
than around $a$ $dozen$. Naturally, the eigenvalues
$\Omega_{\nu}$ and eigenstates
$|\nu>$ are different in each case. So, the strength, that in RPA was
distributed over hundreds or thousands levels, in WFM is concentrated
only on several levels, i.e.
averaging of levels is accompanied by the redistribution of the
strength. The variable $\delta R_{\lambda\mu}^{\tau}(t)$ is the
quantum mechanical observable, so its value should not depend on the
basis $|\nu>$. Hence,
the right hand sides of (\ref{strength}), calculated by two methods,
should coincide if both methods are mutually consistent. This
statement can e.g. be checked with the help of sum
rules. Generally in RPA sum rules are well fulfilled for a
sufficiently large particle hole space which in realistic cases can
become quite significant, whereas in WFM sum rules are generally
already well fulfilled even  with a small number of low rank moments
(see e.g.\cite{Bal,Piper,Duran,Sinich}).

 The essential difference between WFM and RPA
methods lies in their practical use. The RPA
equations (\ref{RPA}) are constructed in such a way that the increase
in dimension does not cause any formal problems and finally it is
only a question of computer power what dimension can be handled.
Quite on the contrary the increase of dimension in WFM is a
nontrivial task. Beyond a certain order of the moments even the
reduction of Cartesian tensors to the irreducible ones becomes a very
difficult task.
However, the spirit of WFM is rather to reproduce the gross structure
of a couple of prominent collective states, a situation which it can
handle very efficiently. However, in cases where there is strong
fragmentation, direct diagonalisation of the standard RPA equations
is more efficient.

In conclusion WFM and RPA are equivalent when the full particle hole
configuration space in RPA and the infinite number of moments in WFM
are considered. However, under truncation of the spaces both methods
have different convergence properties. In the general case for WFM
only a few moments are sufficient to get the correct gross structure
of the collective part of the spectrum, whereas in RPA one in general
 must take into account a
quite large configuration space to produce reasonable results.

\section{Green's function method}

One of the important subjects of comparing RPA and WFM methods
are the current distributions. The WFM
method, a priori, can not give the exact results, because it
deals only with integrals over the whole phase space.
It would therefore be very interesting to evaluate
the accuracy of this approximation by comparing it with the exact result.
 Unfortunately, even for the simple model HO+QQ it is
impossible to derive in RPA closed analytical expressions for
currents of the scissors mode and IVGQR. That is why we
consider in this section the Green's Function (GF) method, which
allows one to find explicit expressions for the currents directly.

Following the paper of H. Kohl, P. Schuck and S. Stringari \cite{Kohl}
we will consider at first the isoscalar case. Conserving on the right
hand side of equation (\ref{fsin}) only the first term of the
sin-function expansion leads to the
Vlasov equation
\begin{equation}
\label{Vlasov}
\frac{\partial f}{\partial t}=
\nabla H_W \cdot \nabla^p f - \nabla^p H_W \cdot \nabla f.
\end{equation}
In our case the Wigner transform  $H_W$ coincides with the classical
Hamiltonian $H_c$.
Having in mind small amplitude vibrations we have to linearize
(\ref{Vlasov}): $f=f_0+f_1 \,$, $ H_c=H_0+H_1,$ with
$f_0$ being the solution of the time independent equation.
The linearized version of (\ref{Vlasov}) is
\begin{equation}
\label{linVlas}
\frac{\partial f_1}{\partial t}
+ \nabla^p H_0 \cdot \nabla f_1
-\nabla H_0 \cdot \nabla^p f_1
=S(\br,\bp,t),
\end{equation}
where $S(\br,\bp,t)=
\nabla H_1 \cdot \nabla^p f_0$.
This equation will be solved with the Green's function method. We have
\begin{equation}
\label{Greq}
(\frac{\partial}{\partial t}
+ \nabla^p H_0 \cdot \nabla
-\nabla H_0 \cdot \nabla^p)G^{(t-t')}(\br\bp,\br'\bp')=\delta(\br-\br')
\delta(\bp-\bp')\delta(t-t')
\end{equation}
with \cite{Kohl}
$$
G^{(t-t')}(\br\bp,\br'\bp')=\delta[\br_c(\br,\bp,t'-t)-\br']
\delta[\bp_c(\br,\bp,t'-t)-\bp']\theta(t-t'),
$$
where $\br_c(\br,\bp,t'-t),\,\bp_c(\br,\bp,t'-t)$ are solutions of
classical equations of motion with initial conditions
$\br,\,\bp$. The solution of (\ref{linVlas}) can be written as
\begin{eqnarray}
f_1(\br,\bp,t)&=&f^h_1+\int\limits_{-\infty}^{\infty}dt'\int d^3r'd^3p'
G^{(t-t')}(\br\bp,\br'\bp')S(\br',\bp',t')
\nonumber\\
&=&f_1^h+\int\limits_{-\infty}^{t}dt'S(\br_c,\bp_c,t'),
\label{f1}
\end{eqnarray}
where $f_1^h$ is the solution of the homogeneous equation. It is
obvious that any function of variables $\br_c$ and $\bp_c$ satisfies
the homogeneous equation, however it does not play any role at
resonance and we therefore will omit it in the forthcoming.

We consider the axially deformed harmonic oscillator with the
quadrupole--quadrupole residual interaction $V_{res}$. Therefore the
single-particle Hamiltonian is
$$H_0=\frac{p^2}{2m}+\frac{m}{2}[\omega_x^2(x^2+y^2)+\omega_z^2z^2]$$
and $H_1=V_{res}$.

We are interested in the part of the residual interaction with
$|\mu|=1$. In accordance with formula (A.1) it can be written as
$$V_{res}=-\kappa_0[Q_{21}(t)q_{2-1}(\br)
+Q_{2-1}(t)q_{21}(\br)]=12\kappa_0 Q_1(t)[xz+yz]$$
with
$$Q_1(t)=2\int\! d\{\bp,\br\}f(\br,\bp,t)xz=
2\int\! d\{\bp,\br\}
f(\br,\bp,t)yz$$
$$=
2\int\! d\{\bp,\br\}
[f_0(\br,\bp)+f_1(\br,\bp,t)]xz=
2\int\! d\{\bp,\br\}
f_1(\br,\bp,t)xz.$$
With the help of the Thomas-Fermi approximation for the static
distribution function
$$f_0=\theta(\epsilon_F-H_0)$$
the right hand side of (\ref{linVlas}) is found to be
$$S(\br,\bp,t)=-12\frac{\kappa_0}{m}Q_1(t)\delta(\epsilon_F-H_0)
[p_xz+p_zx+p_yz+p_zy].$$

The classical trajectories are determined by the solution of Hamilton
equations\\
$\dot r_{c,i}=\frac{\partial H_0}{\partial p_i}$,
$\,\dot p_{c,i}=-\frac{\partial H_0}{\partial r_i}$ with $i=x,y,z.$
In our case they are
$$r_{c,i}(t)=r_i\cos\omega_it+\frac{p_i}{m\omega_i}\sin\omega_it,
\quad p_{c,i}(t)=p_i\cos\omega_it-m\omega_ir_i\sin\omega_it.$$
Formula (\ref{f1}) then gives
\begin{eqnarray}
f_1(\br,\bp,t)&=&-6\frac{\kappa_0}{m}\delta(\epsilon_F-H_0)
\int\limits_{-\infty}^tdt'Q_1(t')\times
\nonumber\\
&&\times\{\frac{1}{\omega_x}(p_x+p_y)z
[\omega_+\cos\omega_+(t'-t)+\omega_-\cos\omega_-(t'-t)]
\nonumber\\
&&+\frac{1}{\omega_z}p_z(x+y)
[\omega_+\cos\omega_+(t'-t)-\omega_-\cos\omega_-(t'-t)]
\nonumber\\
&&+\frac{1}{m\omega_x\omega_z}(p_x+p_y)p_z
[\omega_+\sin\omega_+(t'-t)-\omega_-\sin\omega_-(t'-t)]
\nonumber\\
&&-m(x+y)z
[\omega_+\sin\omega_+(t'-t)+\omega_-\sin\omega_-(t'-t)]\},
\label{f11}
\end{eqnarray}
where $\omega_{\pm}=\omega_x\pm\omega_z.$

So, we have derived a complicated integral equation for the
perturbed distribution function which may not easily be solved in
general. As a matter of fact the analytic possibilities of the
Green's function method
are, without further consideration, exhausted at this point. However,
expressions (\ref{Greq}) and (\ref{f1}) point to the possibility to
use the so-called pseudo particle method \cite{Gupta}, in case the
classical trajectories are not known analytically.

In order to proceed to the evaluation of the eigenfrequencies and
transition probabilities we again apply the method of moments.
Integrating (\ref{f11}) over the whole phase space with the weights
$xz,\,p_xp_z,\,zp_x+xp_z$ and $zp_x-xp_z$ we obtain the following set
of coupled integral equations
\begin{eqnarray}
Q_1(t)&=&\beta\int\limits_{-\infty}^tdt'Q_1(t')
[\omega_+\sin\omega_+(t'-t)+\omega_-\sin\omega_-(t'-t)],
\nonumber\\
P_1(t)&=&-\beta m^2\omega_x\omega_z\int\limits_{-\infty}^tdt'Q_1(t')
[\omega_+\sin\omega_+(t'-t)-\omega_-\sin\omega_-(t'-t)],
\nonumber\\
L_1(t)&=&-\beta m\int\limits_{-\infty}^tdt'Q_1(t')
[\omega_+^2\cos\omega_+(t'-t)+\omega_-^2\cos\omega_-(t'-t)],
\nonumber\\
I_y(t)&=&-\beta m\omega_+\omega_-\int\limits_{-\infty}^tdt'Q_1(t')
[\cos\omega_+(t'-t)+\cos\omega_-(t'-t)],
\label{moment}
\end{eqnarray}
where
\begin{eqnarray}
\beta&=&\frac{2\kappa_0\pi^3\epsilon_F^4}
{m^2\omega_x^4\omega_z^3}\frac{4}{(2\pi\hbar)^3}
\nonumber\\
&=&12\kappa_0\int\! d\{\bp,\br\}x^2z^2\delta(\epsilon_F-H_0)=
\frac{12\kappa_0}{m^4\omega_x^2\omega_z^2}
\int\! d\{\bp,\br\}p_x^2p_z^2\delta(\epsilon_F-H_0)
\nonumber\\
&=&\frac{12\kappa_0}{m^2\omega_z^2}
\int\! d\{\bp,\br\}x^2p_z^2\delta(\epsilon_F-H_0)=
\frac{12\kappa_0}{m^2\omega_x^2}
\int\! d\{\bp,\br\}z^2p_x^2\delta(\epsilon_F-H_0)
\nonumber
\end{eqnarray}
and the following notation is introduced
$$P_1(t)=2\int\! d\{\bp,\br\}f_1(\br,\bp,t)p_xp_z,\quad
L_1(t)=2\int\! d\{\bp,\br\}f_1(\br,\bp,t)(zp_x+xp_z),$$
$$I_y(t)=2\int\! d\{\bp,\br\}f_1(\br,\bp,t)(zp_x-xp_z).$$
By simple means these equations are reduced to a set of
differential equations. At first we perform time derivatives of all
equations in (\ref{moment}):
\begin{eqnarray}
\dot Q_1(t)&=&-\beta\int\limits_{-\infty}^tdt'Q_1(t')
[\omega_+^2\cos\omega_+(t'-t)+\omega_-^2\cos\omega_-(t'-t)],
\nonumber\\
\dot P_1(t)&=&\beta m^2\omega_x\omega_z\int\limits_{-\infty}^tdt'Q_1(t')
[\omega_+^2\cos\omega_+(t'-t)-\omega_-^2\cos\omega_-(t'-t)],
\nonumber\\
\dot L_1(t)&=&-\beta m\left\{(\omega_+^2+\omega_-^2)Q_1(t)+
\int\limits_{-\infty}^tdt'Q_1(t')
[\omega_+^3\sin\omega_+(t'-t)+\omega_-^3\sin\omega_-(t'-t)]\right\},
\nonumber\\
\dot I_y(t)&=&-\beta m\omega_+\omega_-\left\{2Q_1(t)+
\int\limits_{-\infty}^tdt'Q_1(t')
[\omega_+\sin\omega_+(t'-t)+\omega_-\sin\omega_-(t'-t)]\right\}.
\label{tderiv}
\end{eqnarray}
Solving (\ref{moment}) with respect of four time integrals (containing
$\sin\omega_{\pm}$ and $\cos\omega_{\pm})$ we can substitute found
expressions into (\ref{tderiv}). We obtain
\begin{eqnarray}
\dot Q_1(t)&=&\frac{1}{m}L_1(t),
\nonumber\\
\dot L_1(t)&=&-m(2\beta+1)(\omega_x^2+\omega_z^2)Q_1(t)+\frac{2}{m}P_1(t),
\nonumber\\
\dot P_1(t)&=&-\frac{m}{2}[(\omega_x^2+\omega_z^2)L_1(t)-
(\omega_x^2-\omega_z^2)I_y(t)],
\nonumber\\
\dot I_y(t)&=&-m(2\beta+1)\omega_+\omega_-Q_1(t).
\label{dynamic}
\end{eqnarray}
Due to conservation of the angular momentum the right hand side of
the last equation must be equal to zero. So we have the requirement
\begin{equation}
 2\beta+1=0,\quad \mbox{or}\quad \kappa_0=
-\frac{m^2\omega_x^4\omega_z^3
}{4\pi^3\epsilon_F^4}\frac{(2\pi\hbar)^3}{4}.
\label{requir}
\end{equation}
With the help of the relation
\begin{equation}
A\langle r^2\rangle=
2\int\! d\{\bp,\br\}r^2f_0=
2\int\! d\{\bp,\br\}r^2\theta(\epsilon_F-H_0)=
\frac{\pi^3\epsilon_F^4(\omega_x^2+2\omega_z^2)}
{3m\omega_x^4\omega_z^3}\frac{4}{(2\pi\hbar)^3}
\label{r2aver}
\end{equation}
and formulae (A.3) for $\omega_x,\,\omega_z$
the expression for $\kappa_0$ is reduced to
\begin{equation}
\kappa_0=
-\frac{m(\omega_x^2+2\omega_z^2)}{12A\langle r^2\rangle }=
-\frac{m\bar\omega^2}{4A\langle r^2\rangle },
\label{kapp0}
\end{equation}
which is just the familiar expression for the self-consistent value of
the strength constant (see Appendix A). This is a rather interesting
result, because the well known formula is obtained without the usual
self consistency requirement \cite{BM}. As it is known, in the absence
of external fields the angular momentum of any system is conserved.
The
short range interparticle interactions depending on the module of the
interparticle distance $|\br_i-\br_j|$ create the scalar, i.e.
rotational invariant, mean field, which exactly repeats the shape of
the nucleus. When we imitate the mean field by a rotational invariant
function, the angular momentum will be conserved independently of the
shape of this function due to a pure mathematical reason: angular
momentum operator commutes with a scalar field. If we use the non
rotational invariant function (as in our case), mathematics
does not help and the shape of the function becomes important. If the
function does not follow exactly the shape of the system, the latter
will react on this inconsistency as on the external field, that leads
to the nonconservation of an angular momentum. Therefore the
requirement of the angular momentum conservation in this case becomes
equivalent to the requirement of the self consistency. This is seen
very well in the method of moments. Integrating equation
(\ref{linVlas}) over the phase space with the weight $zp_x-xp_z$ we
obtain the dynamical equation for $I_y$
\begin{equation}
\frac{d}{dt}I_y=m(\omega_z^2-\omega_x^2)Q_1+\alpha(\langle x^2\rangle
-\langle z^2\rangle)Q_1.
\label{dinIy}
\end{equation}
The requirement of the angular momentum conservation gives the
following relation
\begin{equation}
m(\omega_z^2-\omega_x^2)=\alpha(\langle z^2\rangle-\langle x^2\rangle).
\label{consis}
\end{equation}
Obviously it is the requirement of the consistency between the shapes
of the potential and the nucleus. In principle this relation is less restrictive
than the standard self consistency requirement \cite{BM}. However, the
latter satisfies equation (\ref{consis}) what can be easily checked
with the help of Appendix A.

So, finally the set of equations (\ref{dynamic}) is reduced to
\begin{eqnarray}
\dot Q_1(t)&=&\frac{1}{m}L_1(t),
\nonumber\\
\dot L_1(t)&=&\frac{2}{m}P_1(t),
\nonumber\\
\dot P_1(t)&=&-m\bar\omega^2[(1+\frac{1}{3}\delta)L_1(t)
-\delta I_y(t)],
\nonumber\\
\dot I_y(t)&=&0.
\label{dynscal}
\end{eqnarray}
Taking into account the relations between the definitions of variables
in (\ref{isosca2}) and (\ref{dynscal})
$$ Q_1=-Re\R_{21},\quad
P_1=-Re\P_{21},\quad
L_1=-2Re\L_{21}$$
(which follow from formulae $r^2_{21}=-z(x+iy)$ and
$(rp)_{21}=-\frac{1}{2}[zp_x+xp_z+i(zp_y+yp_z)]$)
and $I_y=2Re\L_{11}$, it is easy to see, that the last set of equations
is identical to (\ref{isosca2}).

With the help of relations (\ref{moment}) the Wigner function
(\ref{f1}) can be written in terms of the Wigner function moments.
Taking into account equations of motion (\ref{dynscal}) and the
time dependence of variables via $e^{-i\Omega t}$ (which leads to the
equality $I_y=0$) one finds
\begin{eqnarray}
f_1(\br,\bp,t)=\frac{3\kappa_0}{\beta m^2}
\delta(\epsilon_F-H_0)
\left\{
-i\Omega m[\frac{1}{\omega_x^2}(p_x+p_y)z
+\frac{1}{\omega_z^2}p_z(x+y)]
\right.
\nonumber\\
\left.
-(\frac{1}{\omega_x^2}+\frac{1}{\omega_z^2})
(p_x+p_y)p_z
+2m^2z(x+y)
\right\}Q_1(t).
\nonumber
\end{eqnarray}
In the case of $\delta=0$ it reproduces the result of \cite{Kohl}.

Having the Wigner function one can calculate transition probabilities
in the same way as in WFM method.

Let us consider now the problem with two sorts of particles:
neutrons and protons. All variables and parameters acquire isotopic
index $\tau$. The part of the residual interaction with $|\mu|=1$,
in accordance with formula (\ref{potenirr}) becomes
$V_1^{\tau}= Z_1^{\tau}(t)[xz+yz]$ with
$Z_1^{\rm n}(t)=12(\kappa Q_1^{\rm n}+\bar\kappa Q_1^{\rm p}),\,$
$Z_1^{\rm p}(t)=12(\kappa Q_1^{\rm p}+\bar\kappa Q_1^{\rm n})$ and
$Q_1^{\tau}=\int\! d\{\bp,\br\}f^{\tau}_1(\br,\bp,t)xz$. The
expression for the Wigner function is obtained from formula
(\ref{f11}) by changing the factor $6\kappa Q_1(t')$ by $\frac{1}{2}
Z_1^{\tau}(t')$.
The dynamical equations for isovector variables
$\bar Q_1=Q_1^{\rm n}-Q_1^{\rm p},$
$\,\bar P_1=P_1^{\rm n}-P_1^{\rm p},$
$\,\bar L_1=L_1^{\rm n}-L_1^{\rm p},$ and
$\,\bar I_y=I_y^{\rm n}-I_y^{\rm p}$ can be derived (in approximation
(\ref{Apr4}))
exactly in the same way as the equations for isoscalar ones.
As it is expected, they coincide with (\ref{scis1}).

As we see, in the considered simple model all results of WFM method
are identical to that of Green's Function (GF) method. Having in mind
also that both methods generate the same set of dynamical equations
for collective variables (Wigner function moments), one could suspect
their identity. In general, this is not quite true.
The principal difference between the two methods is more or less
obvious. In the GF method one finds first the formal solution of
eq. (\ref{sincyclic})
and only afterwards one takes the phase space moments of the found
Wigner function to obtain the final solution of the physical problem.
In the WFM method one takes from the beginning the phase space moments
of equation (\ref{sincyclic}) without any attempts to find the
``natural" expression for the Wigner function.

The reason of coincidence of all results is quite simple.
For the harmonic oscillator with
multipole--multipole residual interaction of arbitrary rank
(multipolarity) the equations of both methods can be derived without any
approximations -- the interaction of the multipolarity $n$ generates
the set of dynamical equations for tensors (moments) of the rank $n$.
For the GF method this is easily
seen from formula (\ref{f1}). In the case of the WFM method it is
seen very well from the structure of equation (\ref{linVlas}).
When one takes the moments of rank $n$, neither the left hand side
no the right hand side of this equation can generate moments of
rank higher than $n$. The coincidence of results in the case of
$n=3$ was demonstrated in \cite{DiToro}.

The power and simplicity of the GF method are restricted by
the potentials for which the analytical solutions for classical
trajectories are known. In the case of realistic forces the GF
method loses its simplicity and transparency, however the pseudo
particle method \cite{Gupta} can still be applied. The WFM method
does not meet any difficulties and continues to be a convenient and
powerful tool for the description of the collective motion what was
demonstrated by calculations with Skyrme forces \cite{Bal}. For an
illustration of this property of the WFM method, currents are a good
example, because the procedure of their construction with WFM is
general enough to be used for any type of force (see section 6.1
below and \cite{BaSc2}).

\section{Flows}

We are interested in the trajectories of infinitesimal displacements
of neutrons and protons during their vibrational motion, i.e. in the
lines of
currents. The infinitesimal displacements are determined by the
magnitudes and directions of the nucleon velocities $\bu(\br,t)$,
given by
\begin{eqnarray}
m \rho(\br,t) \bu(\br,t)&=&
\int\! \frac{4d^3p}{ (2\pi\hbar)^3}\, \bp f(\br,\bp,t)
\nonumber\\
&=&\frac{4}{(2\pi\hbar)^{3}}
\int\! d^3s\int\! d^3p
\: \bp \exp(- i\bp\cdot
 \bs/\hbar)\rho(\br+\frac{\bs}{2}, \br-\frac{\bs}{2},t)
\nonumber\\
&=&-2i\hbar \{(\nabla-\nabla')
\rho(\br, \br',t)\}_{r=r'}
=-\frac{i\hbar}{2}\sum_{\sigma,\tau} \{(\nabla-\nabla')
\rho(\br\sigma\tau, \br'\sigma\tau,t)\}_{r=r'}
\nonumber\\
&=&-\frac{i\hbar}{2} \sum_{pq}\sum_{\sigma,\tau}
\{\phi_p^*(\br\sigma\tau)\nabla\phi_q(\br\sigma\tau)
 -\phi_q(\br\sigma\tau)\nabla\phi_p^*(\br\sigma\tau)\}
<\Psi|a_p^{\dagger}a_q|\Psi>
\nonumber\\
&=&m\sum_{pq}j_{pq}(\br)\rho_{qp}(t)
=m<\Psi|\sum_{pq}j_{pq}(\br)a_p^{\dagger}a_q|\Psi>
\nonumber\\
&=&m<\Psi|\hat J(\br)|\Psi>.
\label{velos}
\end{eqnarray}
The current density operator $\hat J(\br)$ has the standard quantum
mechanical definition \cite{Ring}:
$$\hat J(\br)
=\sum_{s=1}^A\hat j_s(\br)
=-\frac{i\hbar}{2m}\sum_{s=1}^A
[\delta(\br-\hat\br_s)\nabla_s
+\nabla_s\delta(\br-\hat\br_s)]
=\sum_{pq}j_{pq}(\br)a^{\dagger}_p a_q,$$
$$j_{pq}(\br)=
-\frac{i\hbar}{2m}<p|[\delta(\br-\hat\br)\nabla
+\nabla\delta(\br-\hat\br)]|q>$$
$$=\frac{i\hbar}{2m}\sum_{\sigma,\tau}
[\phi_q(\br\sigma\tau)\nabla\phi^*_p(\br\sigma\tau)
-\phi^*_p(\br\sigma\tau)\nabla\phi_q(\br\sigma\tau)]
=4\frac{i\hbar}{2m}[\phi_q(\br)\nabla\phi^*_p(\br)
-\phi^*_p(\br)\nabla\phi_q(\br)].$$
The variation of $\bu$ generated by the external field (\ref{Exfield})
is
\begin{eqnarray}
\rho^{eq}(\br)\delta \bu(\br,t)
&=&\sum_{pq}j_{pq}(\br)\rho_{qp}^{(1)}(t)
\nonumber\\
&=&\sum_{\nu}[<0|\hat J(\br)|\nu>c_{\nu}
-<\nu|\hat J(\br)|0>
\bar c_{\nu}]e^{-i\Omega t}
\nonumber\\
&&+\sum_{\nu}[<\nu|\hat J(\br)|0>c_{\nu}^*
-<0|\hat J(\br)|\nu>\bar c_{\nu}^*]e^{i\Omega t}.
\label{deltau}
\end{eqnarray}

To proceed further three options are possible.

\subsection{WFM method}

The first way was developed within the WFM approach \cite{Bal}. It
allows one to derive
an approximate analytical expression for $\delta \bu(\br,t)$. The main
idea lies in the parametrization of infinitesimal displacements
${\bf \xi}(\br,t)$. Let us recall the main points. By definition
$\di\delta \bu_i(\br,t)=\frac{\partial \xi_i(\br,t)}{\partial t}$.
The displacement $\xi_i$ is parametrized \cite{BaSc2,Bal}
by the expansion
\begin{equation}
\xi_i(\br,t)=G_i(t)+\sum_{j=1}^3G_{i,j}(t)x_j+\sum_{j,k=1}^3
G_{i,jk}(t)x_jx_k+\sum_{j,k,l=1}^3G_{i,jkl}(t)x_jx_kx_l+\cdots
\label{displ}
\end{equation}
which, in principle, is infinite, however one makes the approximation
keeping only the first terms and neglecting the remainder.
For example, in \cite{BaSc2} only the two first terms were kept.
It turned out, that the $G_i$ do not contribute to the final results
due to the triplanar symmetry of considered nuclei. The coefficients
$G_{i,j}$ were expressed analytically in terms of the variables
$\bar\R_{21}(t)$ and $\bar\L_{11}(t)$. Using the dynamical relations
between $\bar\R_{21}(t)$ and $\bar\L_{11}(t)$
given by the last equation of the set (\ref{scis1}), the final
formulae for $\xi_i(\br,t)$ were found to be
\begin{equation}
\xi_1=\sqrt2 B\bar\J_{13}x_3,\quad
\xi_2=\sqrt2 B\bar\J_{23}x_3,\quad
\xi_3=\sqrt2 A(\bar\J_{13}x_1+\bar\J_{23}x_2)
\label{displac}
\end{equation}
with
$$\bar\J_{13}=(\bar\R_{2-1}-\bar\R_{21})/2,\quad
\bar\J_{23}=i(\bar\R_{2-1}+\bar\R_{21})/2,$$
\begin{eqnarray}
A=\frac{3}{\sqrt2}[1-2\frac{\bar\omega^2}{\Omega^2}(1-\alpha)\delta]
/[Q_{00}(1-\frac{2}{3}\delta)],
\nonumber\\
B=\frac{3}{\sqrt2}[1+2\frac{\bar\omega^2}{\Omega^2}(1-\alpha)\delta]
/[Q_{00}(1+\frac{4}{3}\delta)].
\label{AiB}
\end{eqnarray}
By definition the infinitesimal displacements $\xi_i$ are the
differentials ($\xi_1=dx, \xi_2=dy, \xi_3=dz$). This fact allows one
to construct the differential equations for current fields. For
example, for the current field in the plane $y=0$ we have
\begin{equation}
\frac{dx}{dz}=\frac{B}{A}\frac{z}{x}\quad\longrightarrow \quad
xdx-\frac{B}{A}zdz=0.
\label{curfield}
\end{equation}
Integrating this equation we find
$$x^2+\sigma z^2= const\equiv c\quad \longrightarrow \quad
\frac{x^2}{c}+\frac{z^2}{c/\sigma}=1,$$
where $\sigma=-B/A.$ Depending on the sign of $\sigma$ this curve will
be either an ellipse or a hyperbola. It was shown in \cite{BaSc2} that
the curve is an ellipse for the scissors mode and it is a hyperbola
for IVGQR (see Figs 1,2). It was shown also, that
the real motion of the scissors mode is a mixture of
rotational and irrotational behaviour.
To get a quantitative measure for the contribution of each kind of
motion, it is sufficient to write the displacement $\vec\xi$ as the
superposition of a rotational component with the coefficient $a$ and
an irrotational one with the coefficient $b$ \cite{Zaw}:
$$\vec\xi=a\vec e_x\times\vec r+b\nabla(yz)=a(0,-z,y)+b(0,z,y).$$
Comparing the components $\xi_y=(b-a)z,\,\, \xi_z=(b+a)y$ with $\xi_2,
\,\,\xi_3$ in (\ref{displac}) we find
$$b-a=\sqrt2\bar J_{23}B,\quad b+a=\sqrt2\bar J_{23}A \quad
\longrightarrow \quad a=\eta(1+\sigma),\quad b=\eta(1-\sigma),$$
\begin{figure}
\begin{center}
\epsfig{file=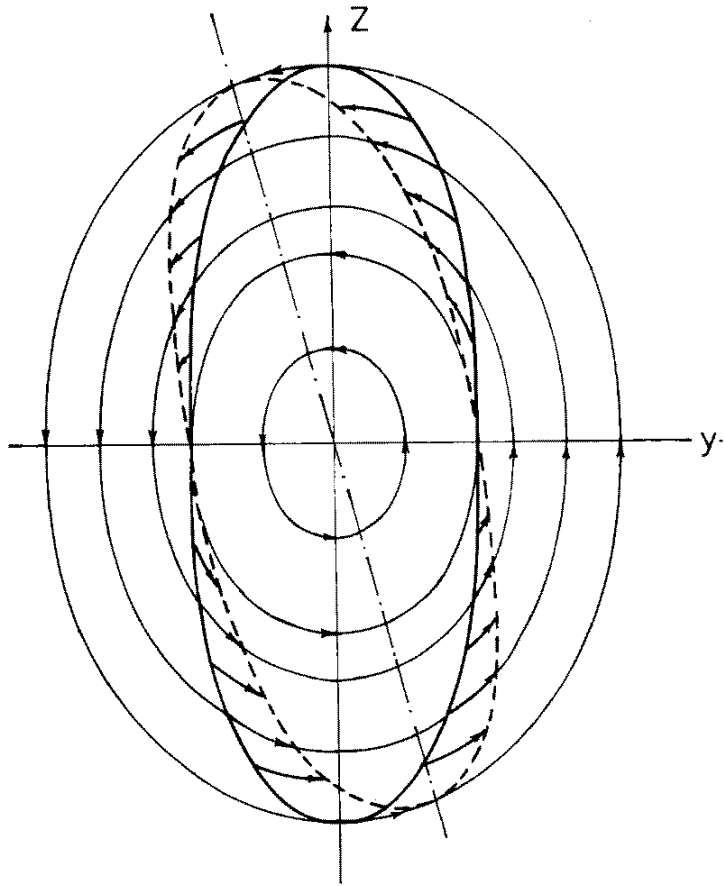,width=10cm}
\end{center}
\caption{Schematic picture of isovector displacements for the
scissors mode. Thin ellipses are the lines of currents. The thick
oval is the initial position of the nucleus' surface (common for protons
and neutrons). The dashed
oval is the final position of the protons' (or neutrons') surface as
a result of an infinitesimal displacements shown by the arrows.\label{fig1}}
\end{figure}
where $\eta=\bar J_{23}A/\sqrt2.$ So, for the scissors mode in the
small $\delta$ limit we have
\begin{equation}
a=2\eta(1-\frac{3}{4}\delta),\quad b=\frac{3}{2}\eta\delta,\quad
b/a\simeq\frac{3}{4}\delta(1+\frac{3}{4}\delta)\approx
\frac{3}{4}\delta,
\label{flowsc}
\end{equation}
i.e. the current of the scissors mode is dominated by rotational
motion. The contributions of the two kinds of motion to the IVGQR are
\begin{equation}
a=\frac{1}{2}\eta\delta,\quad b=2\eta(1-\frac{1}{4}\delta),\quad
a/b\simeq\frac{1}{4}\delta(1+\frac{1}{4}\delta)\approx
\frac{1}{4}\delta,
\label{flowGR}
\end{equation}
i.e. the current of the IVGQR is dominated by irrotational motion.

\begin{figure}
\begin{center}
\epsfig{file=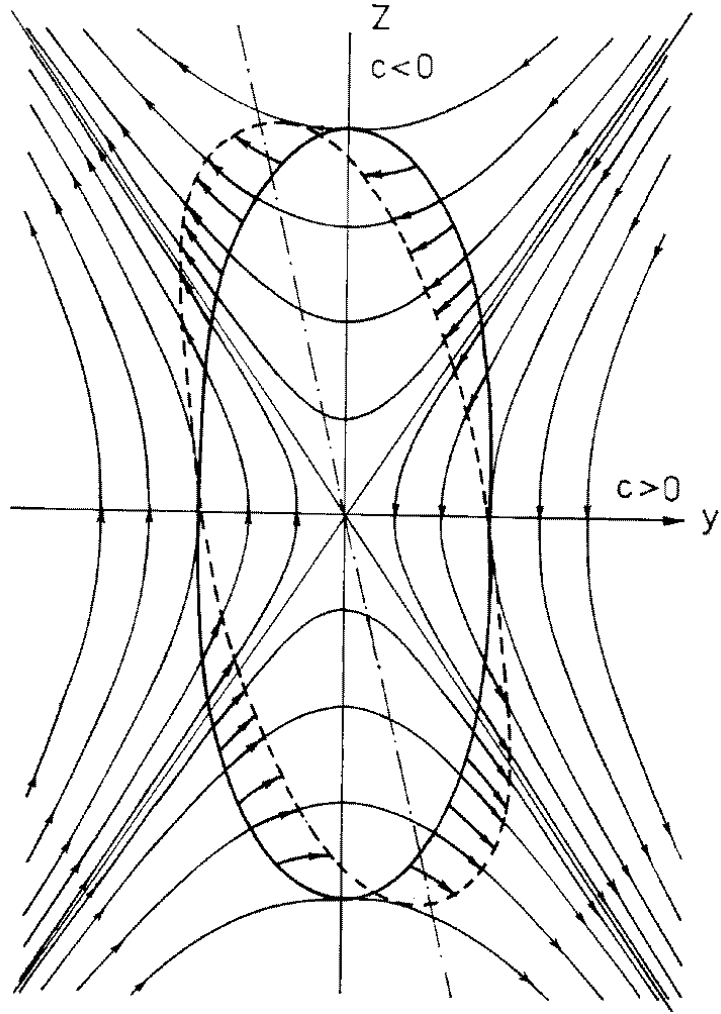,width=10cm}
\end{center}
\caption{Schematic picture of isovector displacements for the
high-lying mode (IVGQR). The lines of currents are shown by thin
lines (hyperbolae). The thick oval is the initial position of the
nucleus' surface (common for protons and neutrons). The dashed
oval is the final position of the protons'
(or neutrons') surface as a result of an infinitesimal displacements
shown by the arrows.\label{fig2}}
\end{figure}
Transitions currents are calculated in WFM analogously to transition
probabilities.
The pole structure of the right hand side of equation (\ref{deltau})
tells us, that the transition current can be calculated by means of
an expression similar to (\ref{Fmatel}):
\begin{equation}
<0|\hat J_i(\br)|\nu>=
\hbar\lim_{\Omega\to\Omega_{\nu}}(\Omega-\Omega_{\nu})\rho^{eq}(\br)
\overline{\dot\xi_i(\br,t)\exp(i\Omega t)}/<\nu|\hat W|0>.
\label{trandis}
\end{equation}
For the  $\xi_i$ from above we obtain (using formulae
(\ref{deltaR}) and (\ref{Cnu}))
\begin{eqnarray}
<0|\hat J_3(\br)|\nu>&=&-i\Omega_{\nu}\rho^{eq}(\br)\frac{A}{\sqrt2}
[<0|\hat R_{2-1}-\hat R_{21}|\nu>x_1
+i<0|\hat R_{2-1}+\hat R_{21}|\nu>x_2],
\nonumber\\
<0|\hat J_2(\br)|\nu>&=&\Omega_{\nu}\rho^{eq}(\br)\frac{B}{\sqrt2}
<0|\hat R_{2-1}+\hat R_{21}|\nu>x_3,
\nonumber\\
<0|\hat J_1(\br)|\nu>&=&-i\Omega_{\nu}\rho^{eq}(\br)\frac{B}{\sqrt2}
<0|\hat R_{2-1}-\hat R_{21}|\nu>x_3.
\label{trancur}
\end{eqnarray}
As it is seen, transition currents are proportional to transition
probabilities.

 If necessary, one can find the next
term of the series (\ref{displ}). To calculate the respective
coefficients $G_{i,jkl}(t)$ in the WFM method
one is obliged to derive (and solve) the set of dynamical equations
for higher (fourth) order moments of the Wigner function.
Examples of similar calculations for third rank tensors can be found
in \cite{JPhysG}.

\subsection{RPA method}

The procedure of constructing the flow distributions in RPA is more
complicated. It is necessary at first to calculate transition currents.
Having solutions (\ref{XY}) for $X_{mi}^{\nu},\,Y_{mi}^{\nu}$ one can
do it with the help of formula
(\ref{matelem}):
\begin{eqnarray}
<0|\hat J(\br)|\nu>
=\sum_{mi}(j_{im}X_{mi}^{\nu}
+j_{mi}Y_{mi}^{\nu})
=K_{\nu}\sum_{mi}\left\{
\frac{j_{im}\Q_{im}^*}{E_{\nu}-\epsilon_{mi}}-
\frac{j_{mi}\Q_{mi}^*}{E_{\nu}+\epsilon_{mi}}
\right\}
\nonumber\\
=K_{\nu}\left\{
\sum_{mi(\Delta N=0)}\left[
\frac{j_{im}\Q_{im}^*}{E_{\nu}-\epsilon_{0}}-
\frac{j_{mi}\Q_{mi}^*}{E_{\nu}+\epsilon_{0}}
\right]+
\sum_{mi(\Delta N=2)}\left[
\frac{j_{im}\Q_{im}^*}{E_{\nu}-\epsilon_{2}}-
\frac{j_{mi}\Q_{mi}^*}{E_{\nu}+\epsilon_{2}}
\right]
\right\}.
\label{trcurr}
\end{eqnarray}
The operator $\Q$ has a finite number of particle hole matrix elements
$\Q_{mi}$, so, in principle, the sums in (\ref{trcurr}) can be
calculated exactly. The same is true for the coefficients $c_{\nu}$
(\ref{Cnu}).
Therefore, in accordance with (\ref{deltau}) one could hope to find
the exact RPA result for the velocity distribution
$\delta \bu(\br,t)$. Unfortunately, because of the pole structure of
coefficients $c_{\nu}(\Omega)$, it can be done for any $\Omega$
except the required frequency $\Omega_{\nu}$ corresponding to the
considered mode (resonance). Of course, it is clear that in the case
of $\Omega$ close enough to $\Omega_{\nu}$ the main contribution into
$\delta \bu$ comes from the single matrix element
$<0|\hat J(\br)|\nu>$.
 That is why, to get an idea about the
distribution of currents in the RPA eigenstate $|\nu>$ it is
sufficient to know the transition matrix element
$<0|\hat J(\br)|\nu>$.
However, even in this simple model one can not find a compact
analytical expression for sums in (\ref{trcurr}) -- the field of
velocities can be constructed only numerically.

As we have already seen it is much more convenient to deal with lines
of currents. The differential equation for them can be derived with
the help of formula (\ref{deltau}). With a e$^{-i\Omega t}$
time dependence we rewrite it in the more convenient form
\begin{eqnarray}
-\rho^{eq}(\br)i\Omega
\xi_i(\br)
=\sum_{\sigma}[<0|\hat J(\br)|\sigma>c_{\sigma}
-<\sigma|\hat J(\br)|0>\bar c_{\sigma}].
\nonumber
\end{eqnarray}
and define the ratio
$$
\frac{\xi_1(\br)}{\xi_3(\br)}=
\frac{\sum_{\sigma}[<0|\hat J_1(\br)|\sigma>c_{\sigma}
-<\sigma|\hat J_1(\br)|0>\bar c_{\sigma}]}
{\sum_{
\sigma}[<0|\hat J_3(\br)|\sigma>c_{\sigma}
-<\sigma|\hat J_3(\br)|0>\bar c_{\sigma}]}.
$$
Remembering the definition of $\xi_i$ and $c_{\sigma}$, multiplying
the numerator and the denominator of the right hand side by
$(\Omega-\Omega_{\nu})$ and taking the
limit $\Omega\to\Omega_{\nu}$ we arrive to the differential equation
\begin{equation}
\frac{dx}{dz}=
\frac{<0|\hat J_1(\br)|\nu>}
{<0|\hat J_3(\br)|\nu>},
\label{currentRPA}
\end{equation}
which determines the lines of currents in the plane $y=const$ for the
resonance state $|\nu>$.

\subsection{Green's function method}

The distribution function being known, one can calculate the
distribution of nuclear currents
$\bj^{\tau}(\br,t)=
m\rho^{\tau}(\br,t)\bu^{\tau}(\br,t)$.
There are no any currents in the equilibrium state, so we have
\begin{eqnarray}
j^{\tau}_x(\br,t)&=&m\rho^{\tau}(\br,t)\delta u^{\tau}_x(\br,t)=
\int\! \frac{2d^3p}{ (2\pi\hbar)^3}\, p_x f^{\tau}_1(\br,\bp,t)
\nonumber\\
&=&-\frac{z}{2m\omega_x}[C_+^{\tau}(t)\omega_++C_-^{\tau}(t)\omega_-]
\int\! \frac{2d^3p}{ (2\pi\hbar)^3}\, p_x^2\delta(\epsilon_F-H_0)
\nonumber\\
&=&-\frac{z}{\omega_x}\frac{2\pi}{3}[2m\epsilon_F-
m^2\omega_x^2(x^2+y^2)-m^2\omega_z^2z^2]^{3/2}
[C_+^{\tau}(t)\omega_++C_-^{\tau}(t)\omega_-]
\nonumber\\
&=&-\frac{z}{2m\omega_x}\rho_0^{\tau}(\br)
[C_+^{\tau}(t)\omega_++C_-^{\tau}(t)\omega_-],
\label{flowx}
\end{eqnarray}
where the following notation is introduced
$$C_{\pm}^{\tau}(t)=\int\limits_{-\infty}^tdt'Z_1^{\tau}(t')
\cos\omega_{\pm}(t'-t).$$
Deriving (\ref{flowx}) we used the approximation (\ref{Apr4})
which means, in particular, that $\omega_i^{\rm n}=\omega_i^{\rm p}$
and $\rho_0^{\rm n}=\rho_0^{\rm p}=\rho_0/2$.
Another component of the flow is
\begin{eqnarray}
j^{\tau}_z(\br,t)=
\int\! \frac{2d^3p}{ (2\pi\hbar)^3}\, p_z f^{\tau}_1(\br,\bp,t)
=-\frac{x+y}{2m\omega_z}\rho_0^{\tau}(\br)
[C_+^{\tau}(t)\omega_+-C_-^{\tau}(t)\omega_-],
\label{flowz}
\end{eqnarray}
With the help of the isovector counterpart of formulae
(\ref{moment}) the functions
$C_{\pm}^{\tau}(t)$ can be written via dynamical variables
$$C_{-}^{\tau}(t)=[L_1^{\tau}(t)
+\frac{\omega_+}{\omega_-}I_y^{\tau}(t)]/\zeta,\quad
C_{+}^{\tau}(t)=-[L_1^{\tau}(t)
+\frac{\omega_-}{\omega_+}I_y^{\tau}(t)]/\zeta$$
with $\di\zeta=\frac{2\pi^3\epsilon_F^4}{3m\omega_x^3\omega_z^2}
\frac{2}{(2\pi\hbar)^3}=\beta m\omega_x\omega_z/6\kappa_0$
and the required combinations are
\begin{eqnarray}
C_{+}^{\tau}(t)\omega_+ +C_{-}^{\tau}(t)\omega_-=
-2\omega_z[L_1^{\tau}(t)-I_y^{\tau}(t)]/\zeta,
\nonumber\\
C_{+}^{\tau}(t)\omega_+ -C_{-}^{\tau}(t)\omega_-=
-2\omega_x[L_1^{\tau}(t)+I_y^{\tau}(t)]/\zeta
\label{combin}
\end{eqnarray}
We are interested in isovector flows
$\bar j_x=j_x^{\rm n}-j_x^{\rm p}$ and
$\bar j_z=j_z^{\rm n}-j_z^{\rm p}$. With the help of the first and
last equations of (\ref{scis1}) we find
\begin{eqnarray}
\bar C_{+}^{\tau}(t)\omega_+ +\bar C_{-}^{\tau}(t)\omega_-=
-2\omega_zi\Omega m[1+2\frac{\bar \omega^2}{\Omega^2}
(1-\alpha)\delta]\bar Q_1/\zeta,
\nonumber\\
\bar C_{+}^{\tau}(t)\omega_+ -\bar C_{-}^{\tau}(t)\omega_-=
-2\omega_xi\Omega m[1-2\frac{\bar \omega^2}{\Omega^2}
(1-\alpha)\delta]\bar Q_1/\zeta.
\label{combvec}
\end{eqnarray}
As a result, we have the explicit expressions for the currents
\begin{eqnarray}
\bar j_z(\br,t)&=&
\frac{i\Omega}{2\zeta}
[1-2\frac{\bar \omega^2}{\Omega^2}(1-\alpha)\delta]\bar Q_1(t)
\rho_0(\br)
\frac{\omega_x}{\omega_z}(x+y),
\nonumber\\
\bar j_x(\br,t)&=&
\frac{i\Omega}{2\zeta}
[1+2\frac{\bar \omega^2}{\Omega^2}(1-\alpha)\delta]\bar Q_1(t)
\rho_0(\br)
\frac{\omega_z}{\omega_x}z.
\label{flows}
\end{eqnarray}
Following the recipe of section 6.1 we can derive the
differential equation for lines of currents, for example,
in the plane $y=0$:
\begin{equation}
\frac{dx}{dz}=\frac{\bar j_x}{\bar j_z}\quad \longrightarrow \quad
\frac{dx}{dz}=
\frac{z}{x}\frac{\omega_z^2}{\omega_x^2}
\frac{1+2\frac{\bar \omega^2}{\Omega^2}(1-\alpha)\delta}
{1-2\frac{\bar \omega^2}{\Omega^2}(1-\alpha)\delta}=
\frac{z}{x}\frac{B}{A}
\label{curgren}
\end{equation}
with $A$ and $B$ defined by (\ref{AiB}). Obviously, this expression
coincides exactly with formula (\ref{curfield}). It is necessary
to emphasize the principal point: the result (\ref{curgren}) is obtained
from the GF method in a direct way, whereas deriving formula
(\ref{curfield})
we made the strong approximation about truncating the expansion
(\ref{displ}) which parametrizes the displacements.
The agreement of both expressions tells us about the internal
consistency of the various approaches to obtain the gross structure
of the flow patterns.

\subsection{Summary of flow calculations}

In conclusion in full RPA one must calculate the currents numerically
leading to fine details (shell effects) whereas in WFM and
GF treatments one obtains their gross structure with
analytical formulas. The latter feature is quite important in order
to understand the real character of the motion under study since
current patterns produced numerically from complicated formulas with
a lot of summations like in (\ref{trcurr}) can hardly be interpreted
physically. A good example is the interplay of the scissors mode and
the isovector giant quadrupole resonance. Looking only at the flow
patterns (see Figs. 1, 2) one would not be able to
tell that the former is mostly rotational with a small amount of an
irrotational component and the other way round for the latter, as
this can be seen from eqs. (\ref{flowsc}, \ref{flowGR}).

\section{Conclusion}

In this paper we made an exhaustive comparison of different methods
to treat collective excitations in nuclei, like the scissors mode,
isovector and isoscalar giant quadrupole resonances. This comparison
was exemplified on the H.O. plus separable
quadrupole--quadrupole force model but it has more general character.

We investigated WFM, RPA and Green's Function (GF) methods. Under
certain circumstances all three methods give essentially the same
results. For example all methods give in our model the same analytical
expressions for energies and transition probabilities for all the
excitations considered. It turned out that
WFM and GF methods are very close to one another. Contrary to RPA,
both work in phase space and incorporate semiclassical aspects, with
no need to introduce a single particle basis. Finally both methods
yield identical sets of dynamical equations for the moments.
However, in the case of realistic forces the GF method loses its
simplicity and the more complicated pseudo particle method has to be
applied, whereas the WFM method continues to be a convenient and
powerful tool for the description of the collective motions as it was
demonstrated in ref. \cite{Bal,Piper,Duran} employing Skyrme forces.

To show the analytical equivalence between WFM and RPA methods one
needs to introduce the dynamical equations for the
transition matrix elements. They can be derived either from the RPA
equations for the amplitudes $X_{kq},Y_{kq}$ or from the WFM dynamical
equations for the moments. This proves the identity of eigenvalues
in both methods under the condition that a complete basis is used in
both cases. However, both methods behave differently when the
dimension of the space is reduced. Actually WFM is designed to use
only rather few moments of low rank. The restricted number of
eigenvalues approximate the collective states in an optimal way,
representing e.g. their centroid positions, as this was shown in
\cite{Bal,Piper,Duran,Providen}. In this sense WFM has similarity
with the sum rule approach \cite{Ring}
which works, however,
only in the cases when practically all strength is exhausted by
one state, whereas WFM method works also in situations when
the strength is distributed among several excitations.
On the contrary in RPA one needs in general a rather large space to
correctly account for the collectivity of e.g. the giant resonances.
At the same time a certain fine structure of the resonances is also
obtained. Concerning the spectrum both methods are thus complementary.
The situation is different for the currents and flow patterns. Since
RPA is a fully quantal approach, the current lines can even in our
simplified model be calculated only numerically. They show a
complicated pattern due to the shell effects. Without further efforts
one will not be able to analyze the nature of the flows. A good
example is given in our model with the low and high lying scissors
mode (the latter being the IVGQR). Due to the analytic formulas found
with WFM and GF methods which naturally lead to smooth current
distributions free of rapidly fluctuating behavior from shell
effects, we were able to show that the low lying scissors mode is
mostly rotational with a
slight admixture of an irrotational component and the other way round
for the IVGQR.

In addition to our earlier work we investigated in detail the so
called synthetic scissors mode which is based on the picture of two
counter rotating proton and neutron mass distributions. Calculating
the overlap of this synthetic scissors mode with the real one we could
show that the squared overlap amounts only to about 60\% in the best
of all cases. We also showed explicitly the orthogonality of the
spurious mode to all other ``intrinsic'' excitations of the model.

Future work in this direction shall be concerned with the scissors
mode in neutron rich nuclei and with the consideration of
superfluidity.

\section*{Appendix A}

It is known that the deformed harmonic oscillator Hamiltonian can be
obtained in a Hartree approximation ``by making the assumption that the
isoscalar part of the QQ force builds the one-body container well"
\cite{Hilt92}. In our case it is obtained quite easily by summing
the expressions for $V^{\rm p}$ and $V^{\rm n}$
(formula (\ref{poten})):
$$V(\br,t)=\frac{1}{2}(V^{\rm p}(\br,t)+V^{\rm n}(\br,t))
=\frac{1}{2}m\,\omega^2r^2+
\kappa_0\sum_{\mu=-2}^{2}(-1)^{\mu}
 Q_{2-\mu}(t)q_{2\mu}(\br).                     \eqno ({\rm A.}1)$$
In the state of equilibrium (i.e., in the absence of an external
field) $Q_{2\pm1}=Q_{2\pm2}=0$. Using the definition \cite{BM}
$Q_{20}=Q_{00}\frac{4}{3}\delta$ and the formula
$q_{20}=2z^2-x^2-y^2$ we obtain the potential of the
anisotropic harmonic oscillator
$$V(\br)=\frac{m}{2}[\omega_x^2(x^2+y^2)+\omega_z^2z^2]$$
with oscillator frequencies
$$\omega_x^2=\omega_y^2=\omega^2(1+\sigma\delta), \quad
\omega_z^2=\omega^2(1-2\sigma\delta),$$
where $\di \sigma=-\kappa_0\frac{8Q_{00}}{3m\omega^2}$. The
definition of the deformation parameter $\delta$ must be reproduced
by the harmonic oscillator wave functions, which allows one to fix
the value of $\sigma$. We have
$$Q_{00}=\frac{\hbar}{m}(\frac{\Sigma_x}{\omega_x}
+\frac{\Sigma_y}{\omega_y}+\frac{\Sigma_z}{\omega_z}),\quad
Q_{20}=2\frac{\hbar}{m}(\frac{\Sigma_z}{\omega_z}
-\frac{\Sigma_x}{\omega_x}),$$
where $\di \Sigma_x=\Sigma_{i=1}^A(n_x+\frac{1}{2})_i$ and $n_x$
is the oscillator quantum number.
Using the self-consistency condition \cite{BM}
$$\Sigma_x\omega_x=\Sigma_y\omega_y=\Sigma_z\omega_z=\Sigma_0\omega_0,$$
where $\Sigma_0$ and $\omega_0$ are defined in the
spherical case, we get
$$\frac{Q_{20}}{Q_{00}}=2\frac{\omega_x^2-\omega_z^2}
{\omega_x^2+2\omega_z^2}=\frac{2\sigma\delta}{1-\sigma\delta}
=\frac{4}{3}\delta.$$
Solving the last equation with respect to $\sigma$, we find
$$\sigma=\frac{2}{3+2\delta}.                \eqno ({\rm A.}2)$$
Therefore, the oscillator frequences and the strength constant can
be written as
$$\omega_x^2=\omega_y^2=\bar\omega^2(1+\frac{4}{3}\delta), \quad
\omega_z^2=\bar\omega^2(1-\frac{2}{3}\delta), \quad
\kappa_0=-\frac{m\bar\omega^2}{4Q_{00}}      \eqno ({\rm A.}3)$$
with $\bar\omega^2=\omega^2/(1+\frac{2}{3}\delta).$
The condition for volume conservation
$\omega_x\omega_y\omega_z=const=\omega_0^3$ makes $\omega$
$\delta$-dependent
$$\omega^2=\omega_0^2\frac{1+\frac{2}{3}\delta}
{(1+\frac{4}{3}\delta)^{2/3}(1-\frac{2}{3}\delta)^{1/3}}
.$$
So the final expressions for oscillator frequences are
$$\omega_x^2=\omega_y^2=\omega_0^2\left(\frac{1+\frac{4}{3}\delta}
{1-\frac{2}{3}\delta}\right)^{1/3}, \quad
\omega_z^2=\omega_0^2\left(\frac{1-\frac{2}{3}\delta}
{1+\frac{4}{3}\delta}\right)^{2/3}.           \eqno ({\rm A.}4)$$

It is interesting to compare these expressions with the very popular
\cite{BM,Ring} parametrization
$$\omega_x^2=\omega_y^2=\omega'^2(1+\frac{2}{3}\delta'), \quad
\omega_z^2=\omega'^2(1-\frac{4}{3}\delta').$$
The volume conservation condition gives
$$\omega'^2=\frac{\omega_0^2}
{(1+\frac{2}{3}\delta')^{2/3}(1-\frac{4}{3}\delta')^{1/3}},$$
so the final expressions for oscillator frequences are
$$\omega_x^2=\omega_y^2=\omega_0^2\left(\frac{1+\frac{2}{3}\delta'}
{1-\frac{4}{3}\delta'}\right)^{1/3}, \quad
\omega_z^2=\omega_0^2\left(\frac{1-\frac{4}{3}\delta'}
{1+\frac{2}{3}\delta'}\right)^{2/3}.          \eqno ({\rm A.}5)$$
The direct comparison of expressions ({\rm A.}4) and ({\rm A.}5)
allows one to establish the following relation between $\delta$ and
$\delta'$:
$$\delta'=\frac{\delta}{1+2\delta},\quad
\delta=\frac{\delta'}{1-2\delta'}.$$
One more parametrization of oscillator frequences can be
found in the review \cite{Zaw}:
$$\omega_x^2=\omega_y^2=\frac{\omega''^2}{1-\frac{2}{3}\delta''}, \quad
\omega_z^2=\frac{\omega''^2}{1+\frac{4}{3}\delta''}.$$
One has from the volume conservation condition
$$\omega''^2=\omega_0^2(1-\frac{2}{3}\delta'')^{2/3}
(1+\frac{4}{3}\delta'')^{1/3},$$
so the final expressions for oscillator frequences are
$$\omega_x^2=\omega_y^2=\omega_0^2\left(\frac{1+\frac{4}{3}\delta''}
{1-\frac{2}{3}\delta''}\right)^{1/3}, \quad
\omega_z^2=\omega_0^2\left(\frac{1-\frac{2}{3}\delta''}
{1+\frac{4}{3}\delta''}\right)^{2/3},          \eqno ({\rm A.}6)$$
that coincide exactly with ({\rm A.}4), i.e.
$\delta''=\delta$.

It is easy to see that equations ({\rm A.}4) correspond to the
case when the
deformed density $\rho(\br)$ is obtained from the spherical density
$\rho_0(r)$ by the scale transformation \cite{Suzuki}
$$(x,y,z)\rightarrow (xe^{\alpha/2},ye^{\alpha/2},ze^{-\alpha})$$
with
$$e^{\alpha}=\left(\frac{1+\frac{4}{3}\delta}
{1-\frac{2}{3}\delta}\right)^{1/3},\quad \delta=\frac{3}{2}
\frac{e^{3\alpha}-1}{e^{3\alpha}+2},          \eqno ({\rm A.}7)$$
which conserves the volume and does not destroy the self-consistency,
because the density and potential are transformed in the same way.

It is necessary to note that $Q_{00}$ also depends on $\delta$
$$Q_{00}=\frac{\hbar}{m}(\frac{\Sigma_x}{\omega_x}
+\frac{\Sigma_y}{\omega_y}+\frac{\Sigma_z}{\omega_z})=
\frac{\hbar}{m}\Sigma_0\omega_0
(\frac{2}{\omega_x^2}+\frac{1}{\omega_z^2})=
Q_{00}^0\frac{1}
{(1+\frac{4}{3}\delta)^{1/3}(1-\frac{2}{3}\delta)^{2/3}},$$
where $Q_{00}^0=A\frac{3}{5}R^2,\,R=r_0A^{1/3}.$
As a result, the final expression for the strength constant becomes
$$\kappa_0=-\frac{m\omega_0^2}{4Q_{00}^0}
\left(\frac{1-\frac{2}{3}\delta}
{1+\frac{4}{3}\delta}\right)^{1/3}
=-\frac{m\omega_0^2}{4Q_{00}^0}e^{-\alpha},$$
that coincides with the respective result of \cite{Suzuki}.

\section*{Appendix B}

To calculate the sums
 $\di \Q_0=\sum_{mi(\Delta N=0)}|\Q_{mi}|^2$ and
$\di \Q_2=\sum_{mi(\Delta N=2)}|\Q_{mi}|^2$
 we employ the sum-rule techniques
of Suzuki and Rowe \cite{Suzuki}.
The well known harmonic oscillator
relations
$$x\psi_{n_x}=\sqrt{\frac{\hbar}{2m\omega_x}}
(\sqrt{n_x}\psi_{n_x-1}+\sqrt{n_x+1}\psi_{n_x+1}),$$
$$\hat p_x\psi_{n_x}=-i\sqrt{\frac{m\hbar\omega_x}{2}}
(\sqrt{n_x}\psi_{n_x-1}-\sqrt{n_x+1}\psi_{n_x+1}) \eqno ({\rm B.}1)$$
allow us to write
$$xz\psi_{n_x}\psi_{n_z}=\frac{\hbar}{2m\sqrt{\omega_x\omega_z}}
(\sqrt{n_xn_z}\psi_{n_x-1}\psi_{n_z-1}
+\sqrt{(n_x+1)(n_z+1)}\psi_{n_x+1}\psi_{n_z+1}$$
$$+\sqrt{(n_x+1)n_z}\psi_{n_x+1}\psi_{n_z-1}
+\sqrt{n_x(n_z+1)}\psi_{n_x-1}\psi_{n_z+1}),$$
$$\frac{\hat p_x\hat p_z}{m^2\omega_x\omega_z}
\psi_{n_x}\psi_{n_z}
=-\frac{\hbar}{2m\sqrt{\omega_x\omega_z}}
(\sqrt{n_xn_z}\psi_{n_x-1}\psi_{n_z-1}
+\sqrt{(n_x+1)(n_z+1)}\psi_{n_x+1}\psi_{n_z+1}$$
$$-\sqrt{(n_x+1)n_z}\psi_{n_x+1}\psi_{n_z-1}
-\sqrt{n_x(n_z+1)}\psi_{n_x-1}\psi_{n_z+1}).     \eqno ({\rm B.}2)$$
These formulae demonstrate in an obvious way that the operators
$$P_0=\frac{1}{2}(zx+
\frac{1}{m^2\omega_x\omega_z}\hat p_x\hat p_z)
\quad\mbox{and}\quad
P_2=\frac{1}{2}(zx-
\frac{1}{m^2\omega_x\omega_z}\hat p_x\hat p_z)$$
contribute only to the excitation of the
$\Delta N=0$ and $\Delta N=2$ states, respectively.
Following \cite{Suzuki}, we express the $zx$ component of
$r^2Y_{21}=\sqrt{\frac{5}{16\pi}}\Q
=-\sqrt{\frac{15}{8\pi}}z(x+iy)$ as
$$zx=P_0+P_2.$$

 Hence, we have
$$\epsilon_0\sum_{mi(\Delta N=0)}|<0|\sum_{s=1}^Az_sx_s|mi>|^2
=\epsilon_0\sum_{mi}|<0|\sum_{s=1}^AP_0(s)|mi>|^2$$
$$=\frac{1}{2}<0|[\sum_{s=1}^A P_0(s),[H,\sum_{s=1}^A P_0(s)]]|0>,
                         \eqno ({\rm B.}3)$$
where $\epsilon_0=\hbar(\omega_x-\omega_z)$.
 The above commutator is easily evaluated for the Hamiltonian with
the potential ({\rm A.}1), as
$$<0|[\sum_{s=1}^A P_0(s),[H,\sum_{s=1}^A P_0(s)]]|0>=
\frac{\hbar}{2m}\epsilon_0
\left(\frac{<0|\sum_{s=1}^Az_s^2|0>}{\omega_x}
-\frac{<0|\sum_{s=1}^Ax_s^2|0>}{\omega_z}\right).\eqno ({\rm B.}4)$$
Taking into account the axial symmetry and using the definitions
$$Q_{00}=<0|\sum_{s=1}^A(2x_s^2+z_s^2)|0>,\quad
  Q_{20}=2<0|\sum_{s=1}^A(z_s^2-x_s^2)|0>,\quad
Q_{20}=Q_{00}\frac{4}{3}\delta,$$
we transform this expression to
$$<0|[\sum_{s=1}^A P_0(s),[H,\sum_{s=1}^A P_0(s)]]|0>=
\frac{\hbar}{6m}\epsilon_0 Q_{00}
\left(\frac{1+\frac{4}{3}\delta}{\omega_x}
-\frac{1-\frac{2}{3}\delta}{\omega_z}\right).    \eqno ({\rm B.}5)$$
With the help of the self-consistent expressions for
$\omega_x,\,\omega_z$ ({\rm A.}3) one comes to the following result:
$$<0|[\sum_{s=1}^A P_0(s),[H,\sum_{s=1}^A P_0(s)]]|0>=
\frac{Q_{00}}{6m}
\frac{\epsilon_0^2}{\bar\omega^2}=
\frac{\hbar^2}{6m}Q_{00}^0
\left(\frac{\omega_0}{\omega_z}
-\frac{\omega_0}{\omega_x}\right)^2.             \eqno ({\rm B.}6)$$
 By using the fact that the matrix elements for the $zy$ component of
$r^2Y_{21}$ are identical to those for the $zx$ component, because of
axial symmetry, we finally obtain
$$\epsilon_0\sum_{mi(\Delta N=0)}|<0|\sum_{s=1}^Ar^2_sY_{21}|mi>|^2=
\frac{5}{16\pi}\frac{Q_{00}}{m\bar\omega^2}\epsilon_0^2=
\frac{5}{16\pi}\frac{Q_{00}^0}{m}\frac{\epsilon_0^2}{\omega_0^2}
\left(\frac{1+\frac{4}{3}\delta}{1-\frac{2}{3}\delta}\right)^{1/3}.
                         \eqno ({\rm B.}7)$$
By calculating the double commutator for the
$P_2$ operator, we find
$$\epsilon_2\sum_{mi(\Delta N=2)}|<0|\sum_{s=1}^Ar^2_sY_{21}|mi>|^2=
\frac{5}{16\pi}\frac{Q_{00}}{m\bar\omega^2}\epsilon_2^2=
\frac{5}{16\pi}\frac{Q_{00}^0}{m}\frac{\epsilon_2^2}{\omega_0^2}
\left(\frac{1+\frac{4}{3}\delta}{1-\frac{2}{3}\delta}\right)^{1/3},
                         \eqno ({\rm B.}8)$$
where $\epsilon_2=\hbar(\omega_x+\omega_z)$.

We need also the sums $\Q_0^{\tau}$ and $\Q_2^{\tau}$ calculated
separately for neutron and proton systems with the mean fields
$V^{\rm n}$ and $V^{\rm p}$, respectively. The necessary formulae
are easily derivable from the already obtained results.
There are no reasons to require the fulfillment of the
self-consistency conditions for neutrons and protons separately,
so one has to use formula ({\rm B.}5). A trivial change of
notation gives
$$<0|[\sum_{s=1}^Z P_0(s),[H^{\rm p},\sum_{s=1}^Z P_0(s)]]|0>=
\frac{\hbar}{6m}\epsilon_0^{\rm p} Q_{00}^{\rm p}
\left(\frac{1+\frac{4}{3}\delta^{\rm p}}{\omega_x^{\rm p}}
-\frac{1-\frac{2}{3}\delta^{\rm p}}{\omega_z^{\rm p}}\right),
                         \eqno ({\rm B.}9)$$

$$\epsilon_0^{\rm p}\sum_{mi(\Delta N=0)}
|<0|\sum_{s=1}^Zr^2_sY_{21}|mi>|^2=
\frac{5}{16\pi}\frac{\hbar}{m}\epsilon_0^{\rm p} Q_{00}^{\rm p}
\left(\frac{1+\frac{4}{3}\delta^{\rm p}}{\omega_x^{\rm p}}
-\frac{1-\frac{2}{3}\delta^{\rm p}}{\omega_z^{\rm p}}\right),
                         \eqno ({\rm B.}10)$$

$$\epsilon_2^{\rm p}\sum_{mi(\Delta N=2)}
|<0|\sum_{s=1}^Zr^2_sY_{21}|mi>|^2=
\frac{5}{16\pi}\frac{\hbar}{m}\epsilon_2^{\rm p} Q_{00}^{\rm p}
\left(\frac{1+\frac{4}{3}\delta^{\rm p}}{\omega_x^{\rm p}}
+\frac{1-\frac{2}{3}\delta^{\rm p}}{\omega_z^{\rm p}}\right).
                         \eqno ({\rm B.}11)$$
The nontrivial information is contained in oscillator frequences of
the mean fields $V^{\rm p}$ and $V^{\rm n}$ (formula (\ref{poten}))
$$(\omega_x^{\rm p})^2=\omega^2[1-\frac{2}{m\omega^2}
(\kappa Q_{20}^{\rm p}+\bar\kappa Q_{20}^{\rm n})],\quad
(\omega_z^{\rm p})^2=\omega^2[1+\frac{4}{m\omega^2}
(\kappa Q_{20}^{\rm p}+\bar\kappa Q_{20}^{\rm n})],$$
$$(\omega_x^{\rm n})^2=\omega^2[1-\frac{2}{m\omega^2}
(\kappa Q_{20}^{\rm n}+\bar\kappa Q_{20}^{\rm p})],\quad
(\omega_z^{\rm n})^2=\omega^2[1+\frac{4}{m\omega^2}
(\kappa Q_{20}^{\rm n}+\bar\kappa Q_{20}^{\rm p})].
                         \eqno ({\rm B.}12)$$

The above-written formulae can also be used to calculate the analogous
sums for operators containing various combinations of momenta and
coordinates, for example, components of an angular momentum,
tensor products $(r\hat p)_{21}$ and $(\hat p^2)_{21}$.
By definition $\hat I_1=y\hat p_z-z\hat p_y, \quad
\hat I_2=z\hat p_x-x\hat p_z$.
 In accordance with ({\rm B.}1), we have
$$x\hat p_z\psi_{n_x}\psi_{n_z}=-i\frac{\hbar}{2}
\sqrt{\frac{\omega_z}{\omega_x}}
(\sqrt{n_xn_z}\psi_{n_x-1}\psi_{n_z-1}
-\sqrt{(n_x+1)(n_z+1)}\psi_{n_x+1}\psi_{n_z+1}$$
$$+\sqrt{(n_x+1)n_z}\psi_{n_x+1}\psi_{n_z-1}
-\sqrt{n_x(n_z+1)}\psi_{n_x-1}\psi_{n_z+1}).     \eqno ({\rm B.}13)$$
 Therefore,
$$\hat I_2\psi_{n_x}\psi_{n_z}=
i\frac{\hbar}{2}(\sqrt{\frac{\omega_z}{\omega_x}}
-\sqrt{\frac{\omega_x}{\omega_z}})
(\sqrt{n_xn_z}\psi_{n_x-1}\psi_{n_z-1}
-\sqrt{(n_x+1)(n_z+1)}\psi_{n_x+1}\psi_{n_z+1})$$
$$+i\frac{\hbar}{2}(\sqrt{\frac{\omega_z}{\omega_x}}
+\sqrt{\frac{\omega_x}{\omega_z}})
(\sqrt{(n_x+1)n_z}\psi_{n_x+1}\psi_{n_z-1}
-\sqrt{n_x(n_z+1)}\psi_{n_x-1}\psi_{n_z+1}).     \eqno ({\rm B.}14)$$
Having formulae ({\rm B.}2) and ({\rm B.}14),
one derives the following expressions for
matrix elements coupling the ground state with $\Delta N=2$ and
$\Delta N=0$ excitations:
$$<n_x+1,n_z+1|\hat I_2|0>=i\frac{\hbar}{2}
\frac{(\omega_x^2-\omega_z^2)}{\omega_x+\omega_z}
\sqrt{\frac{(n_x+1)(n_z+1)}{\omega_x\omega_z}},$$
$$<n_x+1,n_z-1|\hat I_2|0>=i\frac{\hbar}{2}
\frac{(\omega_x^2-\omega_z^2)}{\omega_x-\omega_z}
\sqrt{\frac{(n_x+1)n_z}{\omega_x\omega_z}},$$
$$<n_x+1,n_z+1|xz|0>=
\frac{\hbar}{2m}\sqrt{\frac{(n_x+1)(n_z+1)}{\omega_x\omega_z}},$$
$$<n_x+1,n_z-1|xz|0>=
\frac{\hbar}{2m}\sqrt{\frac{(n_x+1)n_z}{\omega_x\omega_z}}.
                         \eqno ({\rm B.}15)$$
It is easy to see that
$$<n_x+1,n_z+1|\hat I_2|0>=
im\frac{(\omega_x^2-\omega_z^2)}{\omega_x+\omega_z}<n_x+1,n_z+1|xz|0>,$$
$$<n_x+1,n_z-1|\hat I_2|0>=
im\frac{(\omega_x^2-\omega_z^2)}{\omega_x-\omega_z}<n_x+1,n_z-1|xz|0>.$$
Due to the degeneracy of the model all particle hole excitations with
$\Delta N=2$ have the same energy $\epsilon_2$
and all particle hole excitations with $\Delta N=0$  have the energy
$\epsilon_0$. This fact allows one to join the
last two formulae into one general expression
$$<ph|\hat I_2|0>=i\hbar m\frac{(\omega_x^2-\omega_z^2)}
{\epsilon_{ph}}<ph|xz|0>.$$
Taking into account the axial symmetry we have an analogous
formula for $\hat I_1$:
$$<ph|\hat I_1|0>=-i\hbar m\frac{(\omega_x^2-\omega_z^2)}
{\epsilon_{ph}}<ph|yz|0>.$$
The magnetic transition operator (\ref{Omagn}) is proportional
 to the angular momentum:
$\di\hat f_{1\pm1}=
-\frac{ie}{4mc}\sqrt{\frac{3}{2\pi}}(\hat I_2\mp i\hat I_1)$
Therefore, we can write
$$<ph|\hat f_{1\pm1}|0>=-\frac{e\hbar}{2c\sqrt5}
\frac{(\omega_x^2-\omega_z^2)}{\epsilon_{ph}}<ph|r^2Y_{2\pm1}|0>.
                         \eqno ({\rm B.}16)$$
Similar calculations for the tensor product
$(r\hat p)_{21}
=-\frac{1}{2}[z\hat p_x+x\hat p_z+i(z\hat p_y+y\hat p_z)]$
lead to the following relation:
$$<ph|(r\hat p)_{21}|0>
=i\frac{m}{\hbar}\sqrt{\frac{2\pi}{15}}
\epsilon_{ph}<ph|r^2Y_{2\pm1}|0>
=i\frac{m}{2\hbar}\epsilon_{ph}<ph|r^2_{21}|0>.  \eqno ({\rm B.}17)$$

Two kinds of particle hole matrix elements are obtained from the
second formula of ({\rm B.}2):
      $$<n_x+1,n_z+1|\hat p_x\hat p_z|0>=-\hbar m\omega_x\omega_z
\sqrt{\frac{(n_x+1)(n_z+1)}{2\omega_x2\omega_z}},$$
$$<n_x+1,n_z-1|\hat p_x\hat p_z|0>=\hbar m\omega_x\omega_z
\sqrt{\frac{(n_x+1)n_z}{2\omega_x2\omega_z}}.$$
Simple comparison with ({\rm B.}15) shows that
      $$<n_x+1,n_z+1|\hat p_x\hat p_z|0>=
-m^2\omega_x\omega_z<n_x+1,n_z+1|xz|0>,$$
      $$<n_x+1,n_z-1|\hat p_x\hat p_z|0>=
m^2\omega_x\omega_z<n_x+1,n_z-1|xz|0>.$$
With the help of the obvious relations
        $$2\omega_x\omega_z=
\omega_x^2+\omega_z^2-\epsilon_0^2/\hbar^2,\quad
-2\omega_x\omega_z=\omega_x^2+\omega_z^2-\epsilon_2^2/\hbar^2$$
these two formulae can be joined into one expression
      $$<ph|\hat p_x\hat p_z|0>=\frac{m^2}{2}
(\omega_x^2+\omega_z^2-\epsilon_{ph}^2/\hbar^2)<ph|xz|0>.$$
By definition $\hat p^2_{21}=-\hat p_z(\hat p_x+i\hat p_y)$ and
$\hat r^2_{21}=-z(x+iy)$, hence,
$$<ph|\hat p^2_{21}|0>=\frac{m^2}{2}(\omega_x^2+\omega_z^2
-\epsilon_{ph}^2/\hbar^2)<ph|r^2_{21}|0>.        \eqno ({\rm B.}18)$$

\end{document}